\documentclass[aps,twocolumn,longbibliography,nofootinbib,showpacs,showkeys]{revtex4-2}
\usepackage[latin1,utf8]{inputenc} 
\usepackage{amssymb}
\usepackage{amsfonts} 
\usepackage{amsthm} 
\usepackage{amsmath} 
\usepackage{bm}
\usepackage{epsfig}
\usepackage{hyperref}
\usepackage{color}
\usepackage{bbold}
\usepackage[T1]{fontenc}
\usepackage{young}
\usepackage{youngtab}
\usepackage{xcolor}
\usepackage{braket}
\usepackage{xcolor}

\def\nn{\nonumber}
\def\ic{\mathrm{i}}

\def \bc {\begin{center}}
\def \ec {\end{center}}
\def \bi {\begin{itemize}}
\def \ei {\end{itemize}}

\def \ba {\begin{array}}
\def \ea {\end{array}}

\def \bea {\begin{eqnarray}}
\def \eea {\end{eqnarray}}

\def \be {\begin{equation}}
\def \ee {\end{equation}}

\newcommand{\la}{\langle}
\newcommand{\ra}{\rangle}

\def\tr {\mathrm{tr}}
\def\vk {\bm{k}}
\def\vd {\bm{d}}

\def\vs {\bm{\sigma}}
\def\vt {\bm{\tau}}

\begin{document}

\title{Localization and entanglement characterization of edge states in HgTe quantum wells in a finite strip geometry}

\author{Manuel Calixto}
\email{calixto@ugr.es}
\affiliation{Departamento de Matem\'atica Aplicada, Universidad de Granada, 
	Fuentenueva s/n, 18071 Granada, Spain}
\affiliation{Instituto Carlos I de F\'\i sica Te\'orica y  Computacional (iC1), Universidad de Granada, Fuentenueva s/n, 18071 Granada, Spain}

\author{Octavio Casta\~nos}
\email{ocasta@nucleares.unam.mx}
\affiliation{Instituto de Ciencias Nucleares, Universidad Nacional Aut\'onoma de  Mexico, Apdo. Postal 70-543, 04510, CDMX, Mexico}

\date{\today}

\begin{abstract}
Quantum information measures are proposed to analyze the structure of near-gap electronic states in  HgTe quantum wells in a strip geometry $(x,y)\in (-\infty,\infty)\times [0,L]$ of finite width $L$. This allows us to establish criteria for distinguishing  edge from bulk states in the topological insulator phase, including the transition region and cutoff of the wave number $k_x$ where edge states degenerate with bulk states. Qualitative and quantitative information on the near-gap Hamiltonian eigenstates, obtained by tight-binding calculations, is extracted from localization measures, like the inverse participation ratio (IPR),  entanglement entropies of the reduced density matrix (RDM)  to the spin sector --measuring quantum correlations due to the spin-orbit coupling (SOC)-- and from correlation functions for  a $y$-space partition. The analysis of IPR and entanglement entropies in terms of spin, wave number $k_x$ and position $y$, evidences a spin polarization structure and spatial confinement of near-gap wave functions at the boundaries $y=0,L$ and low $k_x$, as correspond to helical edge states.  IPR localization measures provide momentum $k_x$ cutoffs from which near-gap states are no longer  localized at the boundaries of the sample and become part of the bulk. Below this $k_x$-point cutoff, the entanglement entropy and the spin probabilities of the RDM also  capture the spin polarization structure of edge states and  exhibit a higher variability compared to the relatively low entropy of the bulk state region. For a real-space partition, the edge-state region in momentum space exhibits lower  correlation modulus, but higher correlation arguments, than the bulk-state region. 
\end{abstract}




\maketitle

\section{Introduction}

Distinguishing edge states from bulk states in topological insulators (TI) is crucial for understanding their unique properties. There are several approaches to identify them. One is the energy-momentum dispersion relation where bulk states form (conduction and valence) bands with an energy gap in the insulating case, whereas edge states manifest as gapless states within the bulk energy gap, crossing from valence to conduction bands.  Angle-resolved photoemission spectroscopy techniques \cite{Lv2019} directly observe the band structure and the presence of edge states within the bulk gap. Transport measurements at low temperatures also identify edge-dominated conductance in the topological insulating regime. This conductance is quantized due to their robustness against scattering, specially in 2D systems. Such is the case with  the quantum spin Hall (QSH) effect discussed in this article, where edge states carry spin-polarized currents (opposite spins propagate in opposite directions), whereas bulk states do not typically show spin texture unless there is a significant spin-orbit coupling (SOC) in the bulk bands. Magneto-transport techniques can also often help  to separate edge from bulk contributions since edge states are generally robust against moderate magnetic fields whereas bulk states exhibit conventional Landau level behavior. Perhaps the most significant characteristic of edge states is that they are confined to the boundaries (edges or surfaces) of the material whereas bulk states extend throughout the material and are delocalized. Scanning tunneling microscopy/spectroscopy techniques are able to visualize spatially localized states at the edges or surfaces, like in 2D honeycomb-structured film with tellurium \cite{Jianzhong2024}. 

However, the  distinction between edge and bulk states may be blurred  in some scenarios related with the inhibition of the topological protection and can cause edge states to merge into the bulk states in the following sense. For example, changes in the material's physical conditions, geometries, high levels of disorder and doping,  intense magnetic interactions and impurities, time-reversal symmetry breakdown, non-equilibrium conditions, etc. Here we shall analyze the case of  finite size effects  in nanostructures like mercury telluride-cadmium telluride  semiconductor  quantum wells (QW) in a finite strip geometry of width $L$. The distinctiveness between edge-localized and bulk states becomes disrupted in nanostructures  with sample sizes $L$ comparable to or smaller than the coherence length, when edge states from opposite sides, $y=0, L$, can overlap and hybridize, leading to a gap opening in the edge state spectrum. Therefore, looking at the energy-momentum dispersion relation, the hybridized edge states can no longer be distinguished from bulk states. When this happens, we shall talk about ``near-gap states'' and try to establish a $k$-point cutoff (or at least a transition region in momentum space) where edge state dispersion enters the bulk and is no longer visible.

In addition to the standard  tight-binding method (see e.g. \cite{PhysRevLett.95.146801,KaneMele05,PhysRevB.74.085308,KonigJPSJapan2008,Cheng_2014}),  there are other analytic procedures  in the current literature to study these  finite size effects on edge states in the TI  phase. Quadratic Hamiltonian expansions around $\Gamma$ or Dirac valley points $K$ and $K'$ in the Brillouin zone are used in \cite{ZhouPhysRevLett.101.246807,KonigJPSJapan2008,Shan_2010} to analytically determine  the expressions of the edge-state wavefunctions in a   finite strip geometry. For the analytic approach in Section \ref{analyticedge}, the symmetric choice   $y\in [-L/2,L/2]$ seems to be more (computationally) convenient than $y\in [0,L]$, although both conventions are equivalent.  Replacing  the wavevector component  $k_y$  by a complex number $-\ic \lambda$, edge states display an exponential behavior $\psi(y)\sim \exp(\lambda y)$  to take into account an exponential localization at the boundaries $y=\pm L/2$. When the coherence length $1/\mathrm{Re}(\lambda)$ of edge states is comparable to the strip width $L$, states from opposite edges can overlap, leading to a gap opening in the edge state spectrum. These near-gap  states exhibit localization properties typical of edge-states for low $k_x$  but they become degenerate with the bulk states from a certain $k$-point cutoff onwards where $\lambda$ becomes  imaginary. However, this analysis is local (only valid in a small neighborhood around $\Gamma$ or $K$ points) and has some limitations.

In this paper, we tackle the problem of finite size effects in the HgTe/CdTe semiconductors from a different (quantum information) perspective, including SOC  due to bulk- and structure-inversion asymmetries (resp. BIA and SIA). The introduction of SOC has also been investigated in \cite{Cheng_2014}, where they use the tight-binding method to determine an  exponential decay of the edge energy gap (with oscillations/modulations) with the strip width $L$ and to prove that this gap is not exactly localized at the $\Gamma$ point of the first Brillouin zone. This energy gap is also affected by an external perpendicular electric field, which tunes the Rashba (SIA)  term of the Hamiltonian model  and has nontrivial consequences on the charge conductance. We confirm this behavior for a more general BIA term including extra electron and hole couplings preserving time reversal symmetry. We also pursue 
the identification of topological order through quantum information (QI) measures and concepts like entanglement entropy. QI tools  have also played an important role in the general understanding of quantum phase transitions. Indeed, entanglement  is  at the heart of 
the interplay between quantum information and quantum phases of matter (see e.g.,  \cite{Jiang2012,ZengEntangTPT19}). Some standard literature on entanglement research for topological phases primarily examines real-space partitions, revealing universal information of topological phases \cite{PhysRevLett.96.110404,IngoPeschel_2003}. The entanglement energy spectrum analysis in \cite{PhysRevLett.101.010504}  unveils edge state information through momentum-resolved counting of entanglement energy levels, as captured by the ``edge-entanglement spectrum correspondence'' \cite{PhysRevB.91.125119}  or the Li-Haldane conjecture \cite{Zache2022entanglementQuantum}. 
Signatures of topological phase transitions in higher Landau levels of HgTe/CdTe quantum wells without SOC from an information theory perspective have been reported in \cite{CALIXTO2022128057}. Other localization measures, like the inverse participation measure (IPR), 
has given useful information about the topological phase transition 2D Dirac materials like silicene \cite{silicene3}. In this paper we adopt a less standard spin-orbital bipartition to analyze the structure of near-gap  states in HgTe QWs with SOC under QI concepts like IPR and entanglement entropy, which turn out to be an interesting ``microscope'' to reveal  details of their internal structure, and proposes a criterion for the differentiation between  edge and bulk states near the gap  attending to their IPR.

The organization of the paper is as follows. In Sec. \ref{secham} we briefly discuss the structure of  different HgTe QW Hamiltonian models considered in the literature and their topological phases. In Sec. \ref{secenergy}  we approach the analysis of edge states in a finite strip geometry of width $L$ from two different perspectives: either looking for   analytic localized eigenvectors  of the low energy Hamiltonian, or by numerically solving the tight-binding model after a lattice regularization, comparing both approaches for some of the mentioned models. The first approach gives us a deeper understanding of the qualitative and internal structure of edge states (specially for the easier uncoupled case), but we shall rather follow the  second approach to extract quantitative information, firstly about the spectrum and the dependence of the energy gap on the strip width $L$ and the Rashba coupling $\xi$, and its non-trivial consequences on the charge conductance of edge states and its potential use in the design of a QSH field effect transistor. In Sec. \ref{secedgeloc} we take a closer look to the localization properties of near-gap states as a function of the spin ($s=\pm 1$), the momentum wave vector $k_x\in[-\pi/a,\pi/a]$, with $a$ the lattice constant, and the position $y\in[0,L]$ between the strip boundaries $y=0,L$. This study sheds light on the spin-polarization structure of edge states at the boundaries. The spreading of near-gap states in momentum ($k_x$) and position ($y$) space is analyzed through the  IPR  concept of QI, which allows to propose a criterion for the differentiation between  edge-like  and bulk-like behavior, attending to their localization properties. Finally, in Sec. \ref{secentang} we use the reduced density matrix (RDM) to the spin sector to analyze spin up/down and spin-transfer probability densities of near-gap states as a function of the momentum $k_x$ and position $y$, paying especial attention to extremal values.  This analysis also sheds light on the spin-polarization structure of edge states. The purity of the RDM (or equivalently, the linear entropy) also gives us information about the degree of entanglement between spin and band (electron-hole) sectors. Extremal entanglement values occur for special values of the position $y$ and momentum $k_x$. Other alternative correlation measures are also analyzed, all of them giving equivalent results. Section \ref{ypartsec} is devoted to correlation functions for a real-space partition. Finally, Sec. \ref{secconclu} is devoted to conclusions.

\section{Model Hamiltonian}\label{secham}

Let us introduce the model by first briefly reviewing its origins. The prediction and subsequent experimental verification of the QSH effect came after investigations on the spin Hall  effect  associated to relativistic spin-orbit couplings in which electric currents can generate spin currents or vice versa \cite{RevModPhys.87.1213,RevModPhys.82.1539}. 
The QSH  state is a non-trivial topological state of quantum matter which is invariant under time reversal transformations (see e.g. \cite{annurevQSH} for a review). It has an energy  gap in the bulk, but it has edge states with different spins moving in opposite directions, that is, counter-propagating modes at each edge. These spin currents flow without dissipation on macroscopic scales. Mathematically motivated by an earlier model of Haldane \cite{PhysRevLett.61.2015}, graphene was proposed by Kane \& Mele as a two dimensional (2D) Dirac material  to exhibit this effect \cite{PhysRevLett.95.146801,KaneMele05}, however the spin currents were too small to be measurable. Another proposal made by Bernevig-Hughes-Zhang (BHZ) \cite{BernevigHgTe,PhysRevLett.96.106802}, considering the HgTe/CdTe QWs, was successful and this new QSH state of matter and spin polarization phenomena  were experimentally confirmed  through the observation of ballistic edge channels \cite{doi:10.1126/science.1148047,PhysRevB.77.125319} and by electrical detection  \cite{Brune2012}. The intrinsic QSH effect can be switched on and off and tuned into resonance through the manipulation of the QW width $\ell$, or the bias electric field across the QW \cite{PhysRevLett.100.056602}. Since these pioneering studies,  many low-dimensional quantum spintronic devices based on the spin-polarized transport in HgTe/CdTe QWs, and other non-magnetic semiconductors, have been proposed (see e.g.  \cite{Zhang_2021}). For example, other QWs exhibiting a similar behavior to  HgTe/CdTe are the so-called type-II semiconductors made from InAs/GaSb/AlSb, which have been studied in \cite{Liu_PhysRevLett.100.236601}, where they suggest to use this system to construct a QSH field effect transistor (FET). The QSH phenomenon was extended to 3D topological insulators (TI); see \cite{TI1,TI2,TI3} for text books and  \cite{RevModPhys.82.3045,RevModPhys.83.1057} for standard reviews on TI.  In this case,  surface states arise with high conductivity properties, like the  alloy Bi$_x$ Sb$_{1-x}$, which exhibits 2D conducting surface states. Effective Hamiltonian models have been proposed to describe this surface states of 3D TI \cite{Hsieh2008, doi:10.1126/science.1167733,Shan_2010}. 

Let us see in more detail the original BHZ model \cite{BernevigHgTe,PhysRevLett.96.106802}.

\subsection{The uncoupled case}

Following standard references like \cite{NovikHgTe,BernevigHgTe,doi:10.1126/science.1148047,KonigJPSJapan2008,RevModPhys.83.1057,Franzbook}, edge states in HgTe/CdTe QWs are described by the following  2D four-band effective Dirac Hamiltonian. The original BHZ Hamiltonian is 
\begin{eqnarray}\label{ham0}
H_\mathrm{BHZ}&=&\frac{\sigma_0+\sigma_z}{2} \otimes h_{+1}+\frac{\sigma_0-\sigma_z}{2} \otimes h_{-1}=\begin{pmatrix} h_{+1} &0\\ 0 & h_{-1}\end{pmatrix}
 ,\nn\\[2mm]
h_s(\vk )&=&\epsilon_0(\vk)\tau_0+\vd_s(\vk)\cdot\vt, \; s=\pm 1,
\end{eqnarray}
where $\vs=(\sigma_x,\sigma_y,\sigma_z)$, together with the $2\times 2$ identity matrix $\sigma_0$,  are Pauli matrices acting on the spin basis, and $\vt=(\tau_x,\tau_y,\tau_z)$  acting on the electron-hole subbands. The wavevector is denoted by $\vk=(k_x,k_y)$. The spin  $s=\pm 1$, $2\times 2$ matrix  Hamiltonians  $h_s(\vk )$ are related by $h_{-1}(\vk)=h_{+1}^*(-\vk )$ (temporarily reversed) and they admit an expansion around the center $\Gamma$ of the first Brillouin zone (FBZ) given by \cite{BernevigHgTe},
\be
\epsilon_0(\vk)=\gamma-\delta\vk^2,\quad \vd_s(\vk)=(\alpha s k_x,-\alpha k_y,\mu-\beta\vk^2),
\ee
where $\alpha, \beta, \gamma, \delta$ and $\mu$ are material parameters that depend on the HgTe QW geometry, in particular on the  HgTe layer thickness $\ell$. The parameter $\gamma$ can be disregarded and we shall set it equal to zero in the following. In  Table \ref{tabla} we provide these material parameters for a HgTe layer thickness $\ell=7$~nm. We shall use these values all along the manuscript unless otherwise stated.  

The Hamiltonian $h_{s}$ eigenvalues are
\begin{equation}
E_{\pm}(\vk) = \epsilon_0(\vk) \pm \epsilon(\vk),\; \epsilon(\vk)\equiv\sqrt{ \alpha^2 \vk^2  + (\mu - \beta  \vk^2)^2},
\end{equation}
 where $(+)$ makes reference to conduction/electron and $(-)$ to valence/hole states. The corresponding Hamiltonian $h_{s}$ spinor eigenvectors are
\begin{equation}
\psi_{s,\pm}(\vk) = \left(s\frac{\mu-\beta  \vk^2 \pm\epsilon(\vk)}{\alpha  (k_x-\ic s k_y)},1\right)^T,\label{eigenvechs}
\end{equation}
where $T$ means transpose. Note that the bulk gap is $E_g=E_+(\bm{0})-E_-(\bm{0})=2|\mu|$. It closes at the $\Gamma$ point  $\vk=\bm{0}$ for the critical HgTe layer thickness $\ell_c\simeq 6.3$~nm, at which $\mu$ changes sign. For negative $\mu$, edge states arise as gapless states (for large enough sample sizes) within the bulk energy gap, crossing from the valence band to the conduction band and with different spins moving in opposite directions (see next Section \ref{secenergy}). Edge states are said to be topologically protected by the time reversal symmetry 
\begin{equation}
\Theta = - i (\sigma_y \otimes \tau_0) K,\label{trs}
\end{equation} 
of the Hamiltonian \eqref{ham0} where $K$ means complex conjugation.

The $\mathrm{sign}(\mu)=\mathrm{sign}(\ell_c-\ell)$ of the  mass or gap parameter $\mu$, for a given HgTe layer thickness $\ell$, differentiates between band insulator ($\ell<\ell_c$) and topological insulator ($\ell>\ell_c$) phases.  The QSH phase is associated with a discrete $\mathbb{Z}_2$ topological invariant \cite{PhysRevLett.95.146802}. Actually,  the Thouless-Kohmoto-Nightingale-Nijs (TKNN) formula  provides the Chern-Pontryagin number 
\be
\mathcal{C}_s=\frac{1}{2\pi}\int \int_{\mathrm{FBZ}} d^2\vk  \left(\frac{\partial \hat{\vd}_s(\vk )}{\partial k_x}\times  
\frac{\partial \hat{\vd}_s(\vk )}{\partial k_y}\right)\cdot \hat{\vd}_s(\vk ), \label{TKNN}
\ee
with $\hat{\vd}_s=\vd_s /|\vd_s |$, which gives 
\be\label{Chern}
\mathcal{C}_s=s[\mathrm{sign}(\mu)+\mathrm{sign}(\beta)],
\ee
so that the system undergoes a topological phase transition (TPT) from  normal ($\ell<\ell_c$ or $\mu/\beta<0$) to inverted ($\ell>\ell_c$ or $\mu/\beta>0$) regimes at the critical HgTe layer thickness $\ell_c$. The QSH conductance is $\sigma_\mathrm{QSH}=e^2 \mathcal{C}_s/h$, with $e^2/h$ the conductance quantum.

\subsection{Spin-orbit coupling}

Now we shall introduce spin-orbit coupling (SOC) that connects the spin blocks $h_{\pm 1}$. It is given by the Hamiltonian
\begin{eqnarray}\label{ham01}
H_\mathrm{SOC} &=& H_\mathrm{BIA}+H_\mathrm{SIA},\\
H_\mathrm{BIA}(\vk )&=&\Delta_z(\sigma_y\otimes\tau_y)\nn\\&& +\frac{\Delta_e}{2}(k_x\sigma_x-k_y\sigma_y)\otimes(\tau_0+\tau_z)\nn\\ &&+\frac{\Delta_h}{2}(k_x\sigma_x+k_y\sigma_y)\otimes(\tau_0-\tau_z),\nn\\
H_\mathrm{SIA}(\vk ) &=& \frac{\xi}{2} ( k_x \sigma_y -k_y \sigma_x) \otimes (\tau_0 + \tau_z).
\end{eqnarray}
The spin-orbit interaction  creates a  bulk inversion asymmetry (BIA) and a  structural inversion asymmetry (SIA) term which manifests as a $k$-linear Rashba term proportional to $\xi$ for the electron
band  (see e.g. \cite{Liu_PhysRevLett.100.236601,KonigJPSJapan2008,Rothe_2010,Franzbook}); a finite Rashba term of this type in HgTe QWs requires the presence of a non-zero electric field $\mathcal{E}_z$ in the $z$ direction, so that $\xi\propto e\mathcal{E}_z$, with $e$ the electric charge.  We shall set $\mathcal{E}_z=1$~mV/nm all along the manuscript, except for the discussion of the variation of the charge conductance with $\xi$ towards the end of Sec. \ref{secenergy} and Fig. \ref{Fig:conductanceRashba}.

The spin-orbit interaction $H_\mathrm{SOC}$ 
will be responsible for the  entanglement between spin blocks of $H_\mathrm{BHZ}$ in the total Hamiltonian
\begin{eqnarray}\label{ham1}
 H=H_\mathrm{BHZ}+H_\mathrm{SOC}.
\end{eqnarray}
Notice that we are arranging Hamiltonian basis states as 4-spinor column vectors of the form
\begin{equation}
 \Psi=(\psi_{\uparrow E},\psi_{\uparrow H},\psi_{\downarrow E}, \psi_{\downarrow H})^T,\label{4spinor}
\end{equation}
where $\uparrow,\downarrow$ makes reference to the spin degree of freedom $s=\pm 1$ and $EH$ denotes the electron and hole bands, respectively.

The introduction of $H_\mathrm{SOC}$ preserves the time reversal symmetry of the total Hamiltonian $H$ and therefore does not affect the topological stability of the nontrivial insulator phase already discussed for $H_\mathrm{BHZ}$. We shall set $\ell=7$~nm and we shall analyze the topological insulator phase for the material parameters given in Table \ref{tabla} unless otherwise stated (for example, to enhance some physical behavior, due to finite size effects, we shall occasionally consider other  values of $\Delta_z$). 

\begin{table}
\begin{center}
 \begin{tabular}{|c|c|c|c|}
  \hline
   $\alpha$ & $\beta$& $\delta$& $\mu$ \\  (meV.nm)& (meV.nm${}^2$)& (meV.nm${}^2$)& (meV)\\ 
\hline
 365 & -686& -512 & -10 \\
\hline
   $\Delta_e$ &  $\Delta_h$ & $\Delta_z$& $\xi/|e\mathcal{E}_z|$\\ 
   (meV.nm)& (meV.nm)& (meV) & (nm${}^2$)\\ 
\hline
 -12.8 & 21.1&  1.6 & 15.6\\
\hline
 \end{tabular}
 \end{center}
 \caption{Material parameters values for a HgTe/CdTe inverted QW of HgTe  layer thickness  $\ell=7$~nm  \cite{RevModPhys.83.1057,Franzbook}.} \label{tabla}
\end{table}

 \section{Density distribution and energy gap for edge states in a finite strip geometry}\label{secenergy}
 
 In order to extract qualitative and quantitative information on edge states, we shall report on two different but complementary approaches to the solution of the Hamiltonian eigenvalue problem. 
 
 \subsection{Analytic approach to the solution of the effective continuous 4-band model}\label{analyticedge}

 Following Ref. \cite{ZhouPhysRevLett.101.246807} (see also  \cite{Shan_2010} for 3D Bi${}_2$Se${}_3$ films grown on a  SiC substrate), the general solution for  edge states in a finite strip geometry can be analytically derived  as follows. We chose the boundaries of the sample to be  perpendicular to the $y$-axis. For a Hamiltonian matrix of size $n$, $n$-spinor states $\Psi(y)=\Phi e^{\lambda y}$  are proposed  as solutions to the Schr\"odinger equation $H(\vk)\Psi(y)=E\Psi(y)$, by replacing $k_x\to k$ and $k_y\to -\ic\partial_y$. To have nontrivial solutions, the $2n$-degree secular polynomial equation $\det[H(k,-\ic \lambda)-E]=0$ in $\lambda$ must be satisfied, which gives $2n$  different roots  $\lambda_j=\lambda_j(k,E), j=1,\dots, 2n$. The explicit expressions of them for full ($n=4$)  BIA and SIA couplings are too bulky to be given here. Instead, we provide the simpler expressions
  \begin{eqnarray}
\lambda_\pm^2(k,E)&=&k^2 +\frac{\alpha ^2-2 \beta  \mu -2 \delta  E \pm f(E)}{2 \left(\beta^2-\delta^2\right)},\\
f(E)&=& \sqrt{\alpha ^4-4 \alpha ^2 (\beta  \mu +\delta  E )+4 (\beta  E +\delta  \mu )^2},\nn
\end{eqnarray}
for the $n=2$  uncoupled case with $2\times 2$  Hamiltonian $h_s$ in \eqref{ham0}, which yields $2n=4$ roots $\lambda_1=\lambda_+, \lambda_2=\lambda_-, \lambda_3=-\lambda_+, \lambda_4=-\lambda_-$. 
 
 Next we impose open boundary conditions $\Psi(y=\pm L/2)=0$ to a general solution $\Psi(y)=\sum_{j=1}^{2n}C_je^{\lambda_j y}$ of $H(k,-\ic \partial_y)\Psi(y)=E\Psi(y)$, with arbitrary coefficients $C_j$. Demanding a nontrivial solution, the determinant $Q(k,E)$  of the $2n\times 2n$ coefficient matrix must be zero. Solving $Q(k,E)=0$ for $E$ gives the dispersion relation $E(k)$ for the edge states. For the particular $n=2$ case, the transcendental equation $Q(k,E)=0$ reads
 \bea
&& \frac{\tanh \left(\frac{\lambda_1 L}{2}\right)}{\tanh \left(\frac{\lambda_2 L}{2}\right)}+\frac{\tanh \left(\frac{\lambda_2 L}{2}\right)}{\tanh \left(\frac{\lambda_1 L}{2}\right)}\nonumber\\ && -\frac{\alpha ^2 \left(\lambda_1^2+\lambda_2^2\right)-\left(\beta^2-\delta ^2\right) \left(\lambda_1^2-\lambda_2^2\right)^2}{\alpha ^2 \lambda_1 \lambda_2}=0,
 \eea
 in concordance with the result presented in Ref. \cite{ZhouPhysRevLett.101.246807}.

\begin{figure}
     \centering
     \includegraphics[width=7 cm]{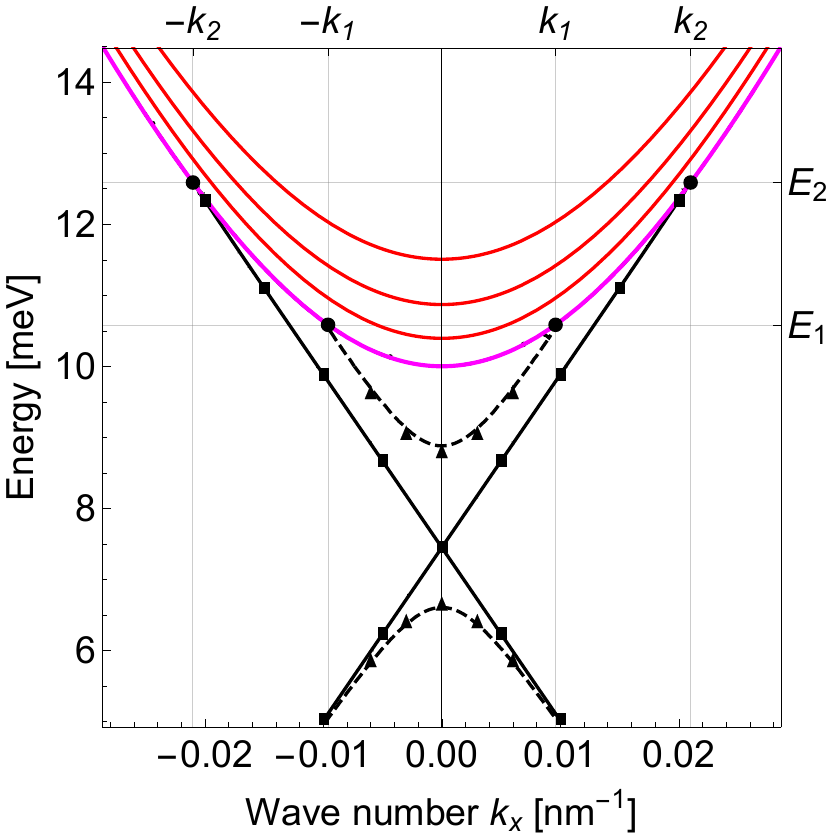}
     \caption{Energy of edge (black) and bulk (red for $k_y\not=0$ and magenta for $k_y=0$) Hamiltonian $h_s$ eigenstates as a function of  $k_x$. Edge and bulk eigenstates merge  at the cutoff wave numbers $\pm k_1=\pm 0.0096$~nm${}^{-1}$ (for strip width $L_1=120$nm, dashed black) and $\pm k_2=\pm 0.021$~nm${}^{-1}$ (for strip width $L_2=500$~nm, solid black), marked by round black dots. The corresponding energies are  $E_1=10.58$~meV and $E_2=12.59$~meV. As a
comparison, results from the numerical diagonalization of tight-binding model of Sec. \ref{tbsec} 
are also plotted as black triangles for $L=120$~nm and black squares for  $L=500$~nm.    }
     \label{fig:comparison0}
 \end{figure}

\begin{figure}
     \centering
     \includegraphics[width=8 cm]{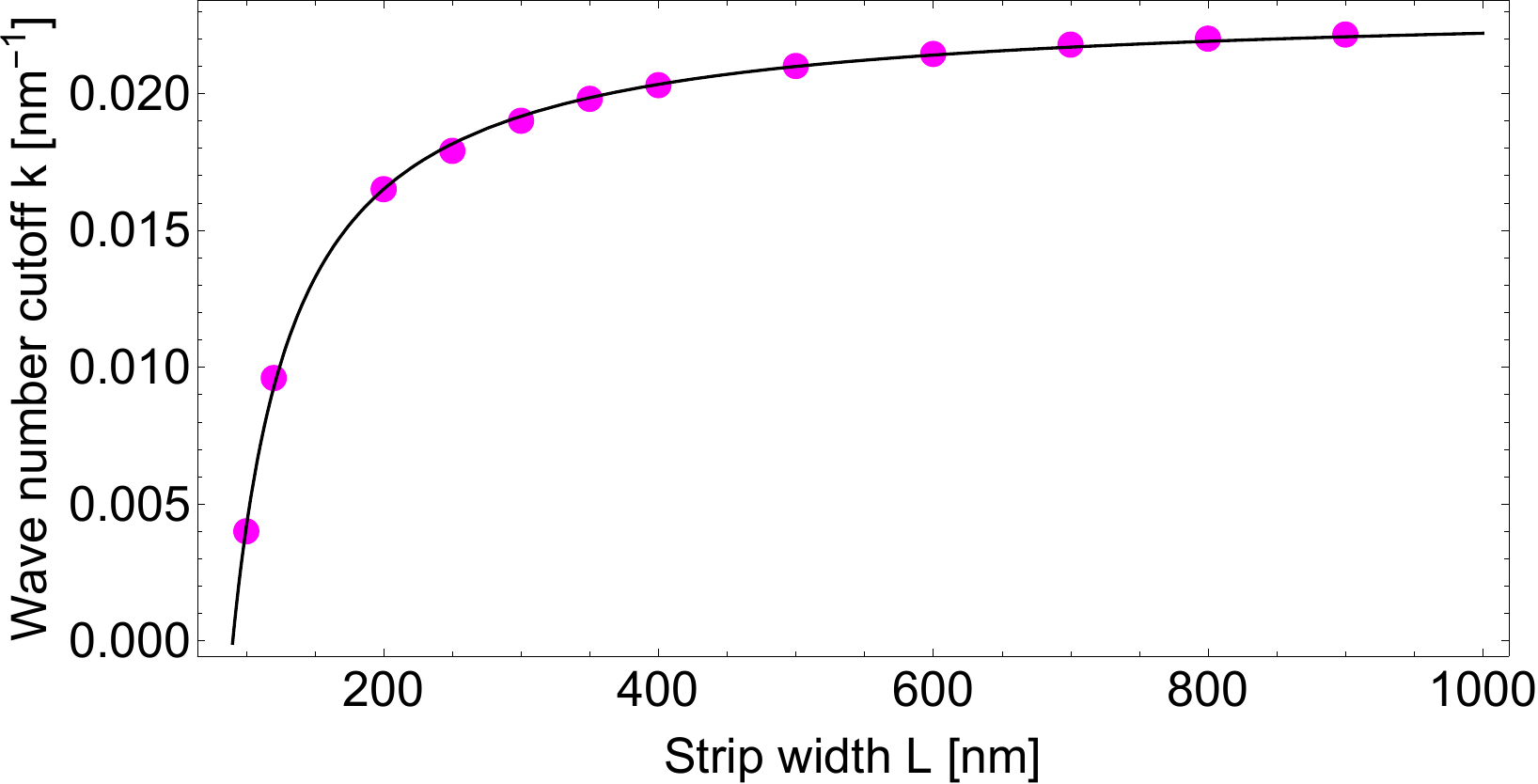}
     \caption{Cutoff wave numbers $k$ (magenta points), at which edge and bulk states merge, for several values  of the strip width  $L\in[100,900]$~nm, for the BHZ Hamiltonian $h_s$. The black curve corresponds to the rational  fitting \eqref{kcutoffL} of these points. }     \label{kcutofffig} 
 \end{figure}

 For the material parameters in Table \ref{tabla}, Figure \ref{fig:comparison0} shows the energy $E(k)$ of the edge states (in black color) of the uncoupled $n=2$ BHZ case for two values, $L_1=120$~nm (dashed) and $L_2=500$~nm (solid), of the strip width. The bulk conduction band  for $h_s$ eigenstates are plotted in red for several values of $k_y$, highlighting the case $k_y=0=\lambda$ in magenta, where edge and bulk eigenstates merge  at the cutoff wave numbers $\pm k_1$ (for strip width $L_1$) and $\pm k_2$ (for strip width $L_2$), marked by round black dots. Edge states show a noticeable energy gap for small $L$ ($E_g\simeq 2.27$~meV for $L=120$~nm) and a linear energy-momentum dispersion relation for large $L$, for which the gap tends to zero (see next paragraph). In Figure \ref{kcutofffig} we plot  the cutoff wave numbers $k$ for several values of the strip width $L\in [100,900]$~nm, together with a rational fitting  of type 
\be
k(L)= 0.0233+\frac{1.0547}{44.906\, -L}\,.\label{kcutoffL}
\ee
 One can infer that there is a limiting cutoff wave number around $k_\infty\simeq 0.0233\,\mathrm{nm}^{-1}$ for large strip widths $L\to\infty$. Note also that $k(L)=0$ gives $L_0\simeq 90$~nm,  which is consistent with the fact that we find no edge solutions of $Q(k,E)=0$ below this value $L_0$, other than those with $\lambda=0$, where edge states become degenerate with bulk states . 
 
The solution of $Q(k,E)=0$ for the coupled case usually suffers from numerical instabilities and we shall prefer the tight-binding approach reported in the next Section \ref{tbsec}.  Tight-binding calculations will  indicate that the introduction of SOC seems to alter the dependence of the cutoff wave numbers $k$ with $L$ reported for the uncoupled case in Eq. \eqref{kcutoffL}.

\begin{figure}
     \centering
     \includegraphics[width=8 cm]{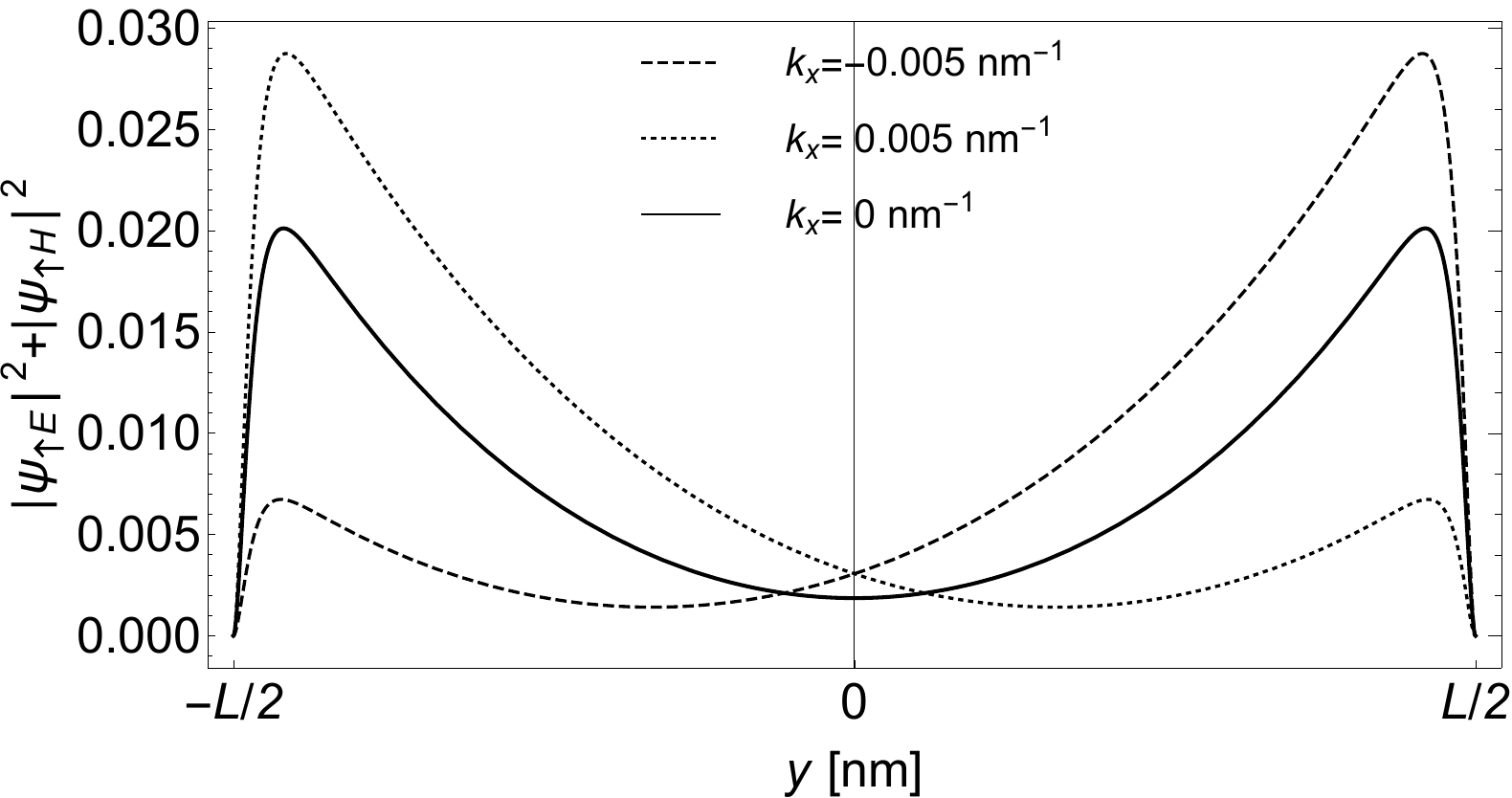}
     \caption{Density distribution $|\psi_{\uparrow E}|^2+|\psi_{\uparrow H}|^2$ of the first (conduction) spin up ($s= 1$)  edge state of the BHZ Hamiltonian $h_s$ for a strip width of $L=120$~nm as a function of $y$. 
     The solid line corresponds to $k_x=0$~nm${}^{-1}$ ($E(0)\simeq 8.88$~meV), the dotted line to $k=0.005$~nm${}^{-1}$ and the dashed line to $k=-0.005$~nm${}^{-1}$ ($E(\pm 0.005)\simeq 9.48$~meV).}
     \label{probdensityfig} 
 \end{figure}

Due to the exponential dependence proposed solution $\Psi(y)\sim e^{\lambda y}$, the real part of $\lambda$ represents the inverse localization length of the edge states, and therefore it  plays a major role in determining whether  the state can be distinguished as a bulk or  edge state, since in finite strip geometries the localization length should be much less than the strip width $L$. In Figure \ref{probdensityfig}  we represent the density distribution of the first analytic solution of the edge state wave function for spin $\uparrow$ as a function of $y$ for three different values of $k_x$ below the cutoff wave number $|k_x|< k_1=0.0096$~nm${}^{-1}$ in Fig. \ref{fig:comparison0} for $L=120$~nm. The results are consistent with those of  Ref. \cite{ZhouPhysRevLett.101.246807}. 
The dominant $\lambda$ root is the one with the larger length scale $|\mathrm{Re}(\lambda)|^{-1}$. The density distribution at the two edges $y=\pm L/2$ is mainly determined by the dominant root. We observe that spin up conduction states are localized at the boundary $y=L/2$ (resp. $y=-L/2$) for negative (resp. positive) $k_x$, the case $k_x=0$ displaying a more balanced configuration. This is a reflect of the already commented spin polarization structure of edge states (see later in Fig.  \ref{Fig:spinpol} for more details).

As proved in Ref. \cite{ZhouPhysRevLett.101.246807}, the edge energy gap $E_g$ shows an exponential decaying $E_g\sim \exp(-\lambda L )$ with $L$. Ref. \cite{Cheng_2014} confirms the exponential decay of $E_g$ with the strip width $L$ but observes an oscillatory behavior as coming from the imaginary part of $\lambda$ and the fact that the gap closes outside the $\Gamma$ point (see next Section \ref{tbsec} for more information). The physical reason for the appearance of a gap opening in the edge state spectrum lies in the fact that edge states from opposite edges couple and hybridize when the width $L$ of the sample is comparable to (or falls below) the coherence length $|\mathrm{Re}(\lambda)|^{-1}$, thus making it  difficult to differentiate between edge and bulk stakes from the energy spectrum. For example, for the Hamiltonian $h_s$ in \eqref{ham0} we have calculated the edge energy gap $E_g^\Gamma=[E_{1c}(0)-E_{1v}(0)]$ at the $\Gamma$ point  for several strip widths between $L_i=120$ and $L_f=500$~nm. A fit of these values provides the expression 
 \begin{equation}
E_g^\Gamma(L)\simeq e^{3.14 - 0.0196 L},\label{gapfit0}
\end{equation}
with  determination coefficient $R^2>0.999$, from which we can derive the  coherence length as  $1/0.019\simeq 51$~nm. This coherence length  gives us an idea of the strip width $L$ below which states from opposite edges couple and hybridize to produce a non trivial (measurable in experiments) energy gap even in the inverted regime (see later in Eq. \eqref{gapfit} for a similar calculation in the  tight-binding approach). Formula \eqref{gapfit0} is in accordance with the analytic approximate expression presented in \cite{ZhouPhysRevLett.101.246807}.

The analytic scheme provides useful information about the internal structure and localization properties  of edge states, the wavenumber region where they exist and the cutoff from which they become part of the bulk. However,  its validity range is limited to a neighborhood of the $\Gamma$ point where the tight-binding Hamiltonian is expanded up to linear or quadratic powers of the wavevector $\vk$ components for a small lattice constant. Moreover, the resolution of the transcendental equations  $Q(k,E)=0$ becomes quite unwieldy for Hamiltonian sizes $n>2$ with coupling, where one has to deal with (sometimes) unavoidable numerical instabilities. 

In the next section we shall deal with the more accurate tight-binding model by  following a numerical approach. This approach will give a better quantitative analysis about the whole (edge and bulk) energy spectrum, although it lacks a clear distinction between edge and bulk states for low $L$, specially in providing a clear wavenumber cutoff beyond which edge states transmute to bulk states. This drawback will be solved by introducing localization and entanglement measures in the next sections.

  \subsection{Lattice regularization and numerical diagonalization of the tight-binding model}\label{tbsec}
  
The general solution for both, bulk and edge, states can be accomplished through a lattice regularization  of the continuum model just replacing 
\begin{equation}k_{x,y}\to a^{-1}\sin(k_{x,y}a),\quad k_{x,y}^2\to 2a^{-2}(1-\cos(k_{x,y}a),
\end{equation}
in the Hamiltonian $H(\vk)$ in \eqref{ham1}, with $a$ the lattice constant. Then, the Brillouin zone (BZ) is $\vk\in(-\pi/a,\pi/a)\times (-\pi/a,\pi/a)$. We shall set $a=2$~nm in all our calculations, in order to compare some of our results with those of Ref. \cite{Cheng_2014}, whose authors also use this value.  Therefore, we shall consider a space discretization of the finite strip along the $y$-direction, with $y_n=na$ and $n=0,\dots, N=L/a$ the site indices, and we shall place hereinafter the lattice boundaries at $y=0, L$, accordingly. The translation between nanometers and lattice units can then be done by simply dividing lengths by  $a=2$.

Following the general procedure of Refs. \cite{KonigJPSJapan2008,Cheng_2014}, one Fourier transforms $k_y$ in the total Hamiltonian $\mathcal{H}=\int_\mathrm{BZ}d\vk H(\vk)c^\dag_{\vk} c_{\vk}$ by substituting the annihilation (viz. creation) operators 
\begin{equation}
 c_{\vk}=\frac{1}{L}\sum_{n=0}^{N} e^{\ic k_y y_n}c_{k,n},\quad y_n=na,\, N=L/a,
\end{equation}
to obtain the tight-binding model Hamiltonian 
\begin{equation}
 \mathcal{H}=\sum_{k,n} \mathcal{E}(k)c_{k,n}^\dag c_{k,n}+\mathcal{T}c_{k,n}^\dag c_{k,n+1}+\mathcal{T}^\dag c_{k,n+1}^\dag c_{k,n},\label{hamTB}
\end{equation}
in position (discrete) $y$ and momentum  $k=k_x$ spaces. 
The $4\times 4$ matrix $ \mathcal{E}(k)$  results from eliminating all terms depending on $k_y$ in the regularized total Hamiltonian $H(\vk)$ in \eqref{ham1}. Those terms then contribute to the matrix  
\begin{equation}
 \mathcal{T}= \left(
\begin{array}{cccc}
 \frac{\beta+\delta }{a^2} & -\frac{\alpha }{2 a} & -\frac{\Delta _e+i \xi}{a} & 0 \\
 \frac{\alpha }{2 a} & \frac{\delta-\beta }{a^2} & 0 & \frac{\Delta _h}{a} \\
 \frac{\Delta _e-i \xi}{a} & 0 & \frac{\beta+\delta }{a^2} & -\frac{\alpha }{2 a} \\
 0 & -\frac{\Delta _h}{a} & \frac{\alpha }{2 a} & \frac{\delta -\beta}{a^2} \\
\end{array}
\right).
 \end{equation}
 The matrix Hamiltonian $\mathcal{H}$ is of size  $4N=4L/a$ and is numerically diagonalized. The Hamiltonian spectrum is composed of both: bulk and edge states.

 Figure \ref{fig:comparison0} shows a comparison of near-gap energies between the analytical and tight-binding approaches for the uncoupled case. These results are consistent with Ref. \cite{ZhouPhysRevLett.101.246807}.

\begin{figure}
	\begin{center}
		\includegraphics[width=8cm]{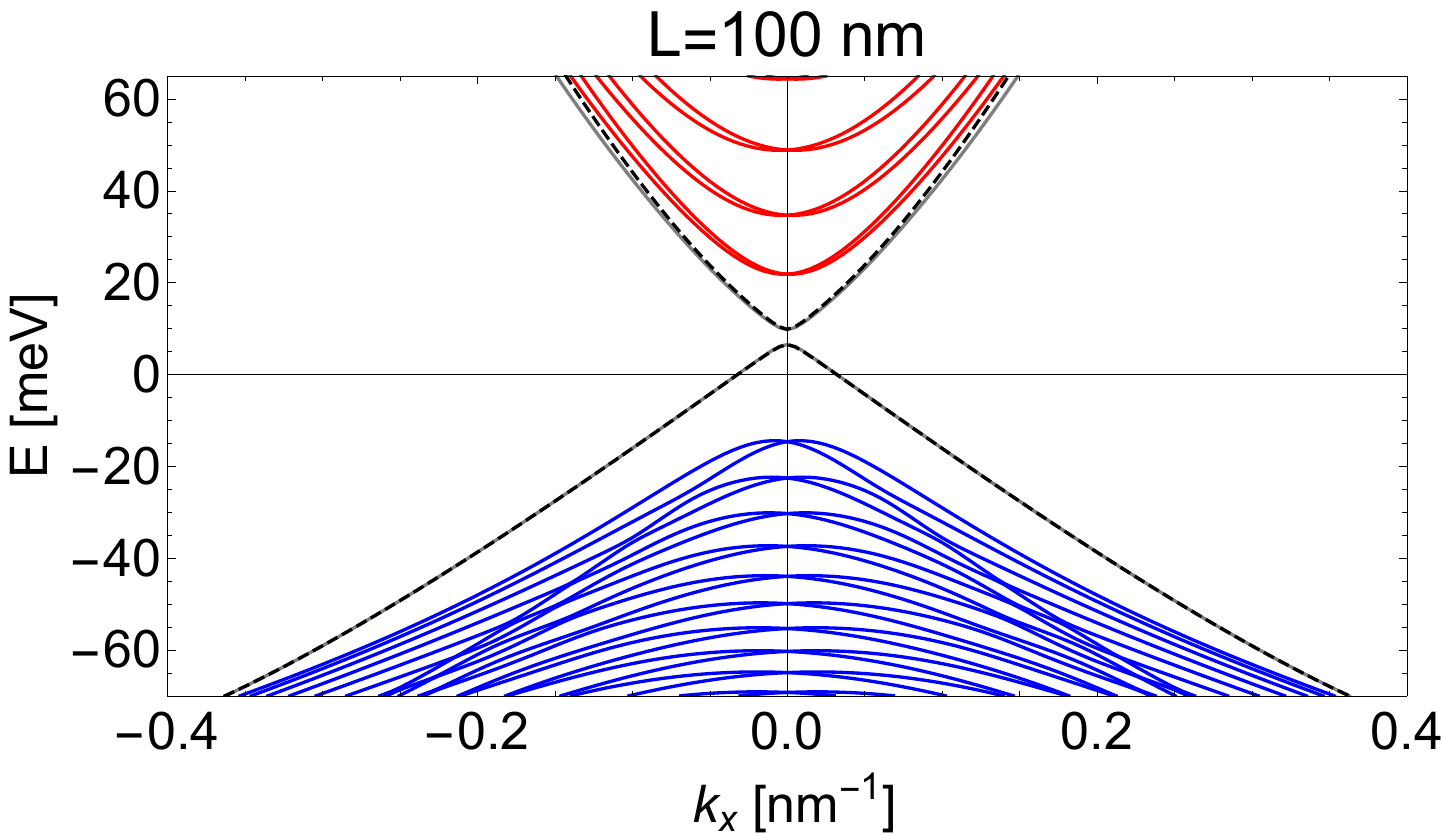}\\ [2mm]
		\includegraphics[width=8cm]{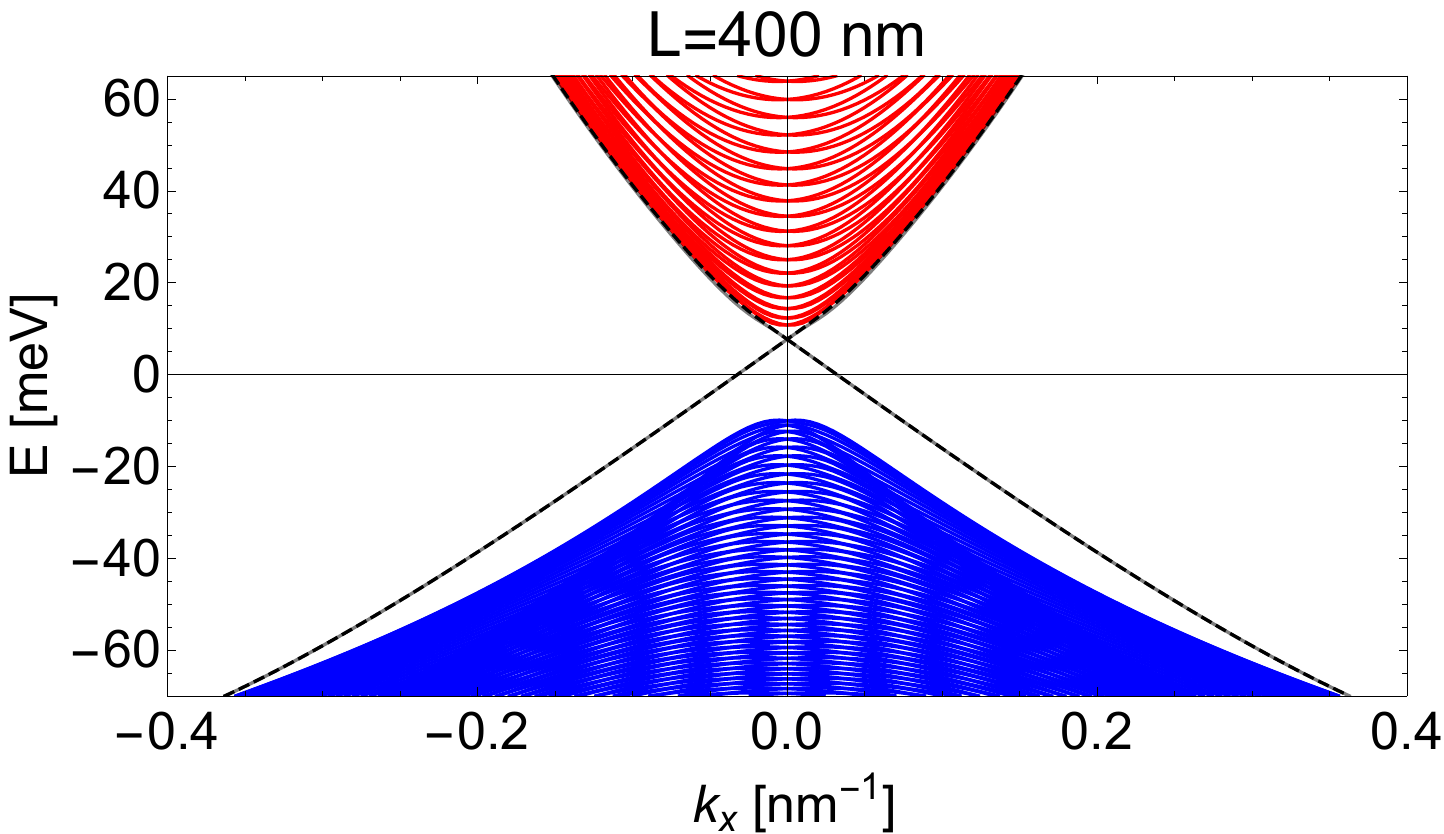}
	\end{center}
	\caption{ Hamiltonian spectra for two strip sizes $L=100$ and $L=400$~nm as  a function of the wavevector component $k_x$ for the material parameters in Table \ref{tabla}. Conduction states in red and valence states in blue color. The four near-gap (``edge to be") states are indicated in black color (solid and dashed). The gap closes as $L$ increases. }
	\label{Fig:Spectrum}
\end{figure}

\begin{figure}
	\begin{center}
		\includegraphics[width=7.5cm]{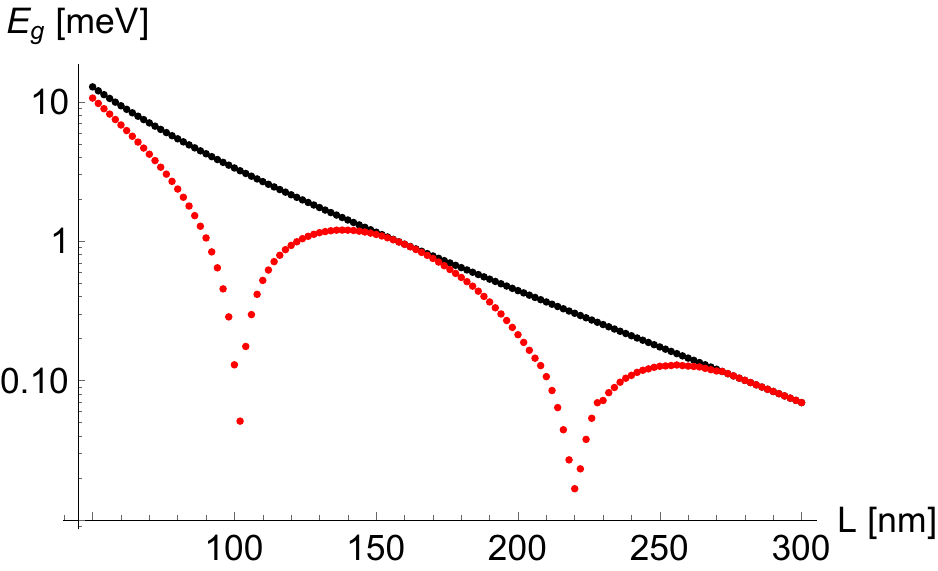}
	\end{center}
	\caption{Energy gap (logarithmic scale) of the  of edge states as a function of the strip width $L$. The gap at the $\Gamma$ point $E_g^\Gamma$ (black color) shows an exponential decay, whereas the minimum gap $E_g$ (red color) exhibits oscillations with sudden drops for some critical strip widths values ($L_c\simeq 100$ and 220~nm). This time  we choose  the SOC parameter $\Delta_z~=~10$~meV for computational convenience. }
	\label{Fig:gaposc}
\end{figure}

 Figure \ref{Fig:Spectrum} shows the energy spectrum $E(k)$ for $L=100$ and $L=400$~nm as a function of the wavevector component $k=k_x$ in the vicinity of the $\Gamma$ point for the Hamiltonian \eqref{hamTB} with full SOC (BIA plus SIA). Bulk conduction/valence $(c/v)$ energy levels $E_{c/v}$ are plotted in red/blue color while the four near-gap (the ``edge to be" for low $k_x$) energy levels, corresponding to the four  4-spinor states denoted by  $\{\Psi_{1c},\Psi_{1v},\Psi_{2c},\Psi_{2v}\}$, are plotted in black color, solid for $\{\Psi_{1c},\Psi_{1v}\}$ and dashed for  $\{\Psi_{2c},\Psi_{2v}\}$. 
Notice that $\Psi_1$ and $\Psi_2$ are nearly degenerated for conduction and valence bands, but the energy $E_{1c}$ is a bit lower than $E_{2c}$ and $E_{1v}$ is slightly higher than $E_{2c}$, so that the energy gap is determined by $E_g=\min_k[E_{1c}(k)-E_{1v}(k)]$, with $k\in(-\pi/a,\pi/a)$. As we already anticipated in Sec. \ref{analyticedge}, this gap shows an exponential decay with modulations/oscillations as function of the strip width $L$, as showed in Figure \ref{Fig:gaposc} (red dots). The value of $k$ minimizing $[[E_{1c}(k)-E_{1v}(k)]$ does not coincide in general with $k=0$ (the $\Gamma$ point) but it is quite close to it. To compute it, one needs a dense mesh of points $k_n$ which requires high computational resources for high $L$. Thankfully, larger values of the SOC coupling $\Delta_z$ require smaller Hamiltonian matrix sizes $N=L/a$ and less computational effort to observe energy gap oscillations. We have chosen $\Delta_z=10$~meV this time (different from the $\Delta_z=1.6$~meV value in Table \ref{tabla} used in Figure \ref{Fig:Spectrum})  just for computational convenience. Sudden gap drops occur at the critical strip widths $L_c\simeq 100$ and $L_c\simeq 220$nm.   The exponential decay of the energy gap  is captured by the gap $E_g^\Gamma=[E_{1c}(0)-E_{1v}(0)]$ at the $\Gamma$ point (black dots in Figure \ref{Fig:gaposc}). A fit of nine values of $E_g^\Gamma$ at $L=100,\dots, 500$, in steps of $\Delta L=50$, provides the expression  
 \begin{equation}
E_g^\Gamma(L)\simeq e^{2.991 - 0.019 L}\label{gapfit}
\end{equation}
with  determination coefficient $R^2>0.999$.  Therefore, for this particular value of  $\Delta_z=10$~meV,  one can derive a coherence length of $1/0.019\simeq 53$~nm,  which is quite similar to the coherence length calculated in \eqref{gapfit0} for the uncoupled case in the analytic approach.

\begin{figure}
	\begin{center}
		\includegraphics[width=4.1cm]{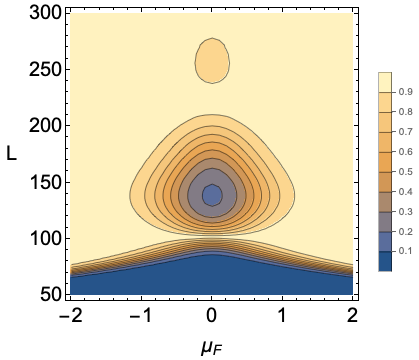}\quad
		\includegraphics[width=4.1cm]{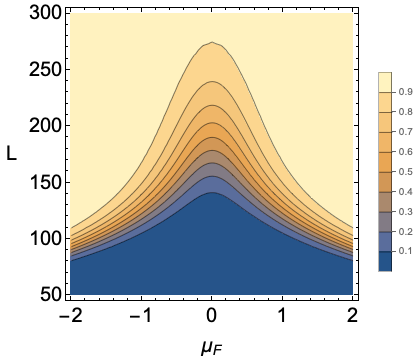}
	\end{center}
\caption{Variation of the conductance (in $2e^2/h$ units)  as a function of the chemical potential $\mu_F$ and the width $L$ of the strip at temperature $T=3$~K. Left panel using the  oscillating gap $E_g$ and right panel using the gap $E_g^\Gamma$ at the $\Gamma$ point (resp. red and black curves of Figure \ref{Fig:gaposc}). Color scale varies from 0 (darkest) to 1 (brightest).}
	\label{Fig:conductance}
\end{figure}

\begin{figure}
	\begin{center}
		\includegraphics[width=4.4cm]{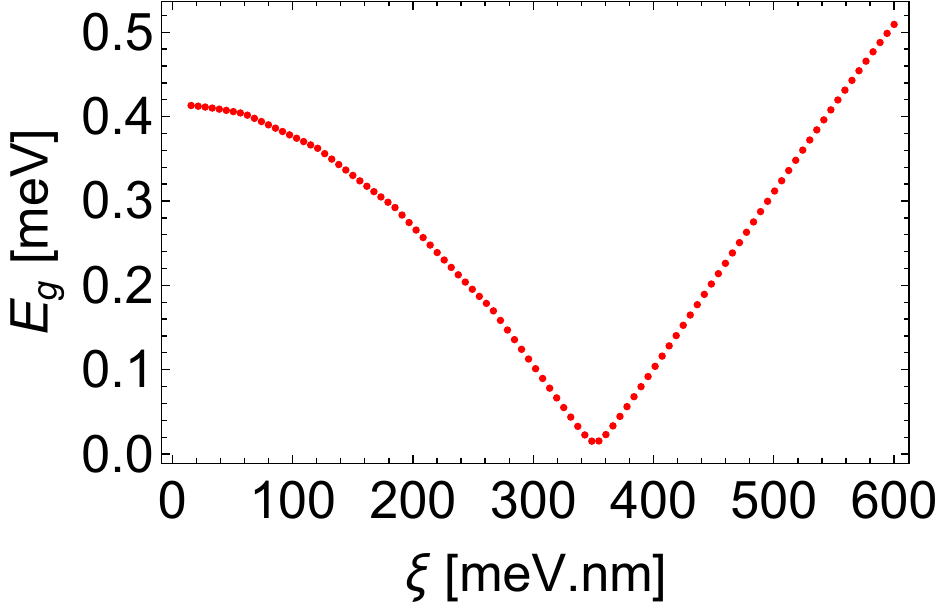}\quad
		\includegraphics[width=3.9cm]{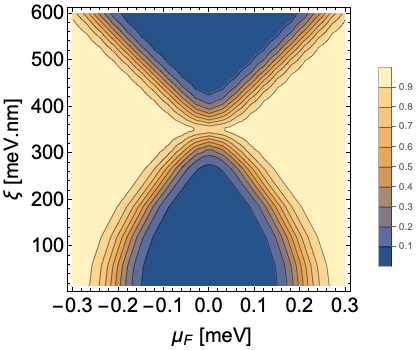}
	\end{center}
\caption{Left panel: energy gap $E_g$ as a function of the Rashba coupling term $\xi$ for $L=200$~nm. Right panel: variation of the conductance (in $2e^2/h$ units)  as a function of the chemical potential $\mu_F$ and $\xi$ for $L=200$~nm and temperature $T=0.3$~K; color scale varies from 0 (darkest) to 1 (brightest).}
	\label{Fig:conductanceRashba}
\end{figure}

These gap oscillations have  non trivial consequences in the charge conductance of the edge states given by the 
Landauer-B\"uttiker formula
\begin{equation}
 G=\frac{1}{e^{(E_g/2-\mu_F)/k_BT}+1}-\frac{1}{e^{(-E_g/2-\mu_F)/k_BT}+1}+1
\end{equation}
in $2e^2/h$ units. In Fig. \ref{Fig:conductance} we plot the charge  conductance  as a function of the chemical potential $\mu_F$ and the width $L$ of the strip at temperature $T=3$~K, for the energy gaps $E_g$ (left panel) and $E_g^\Gamma$ (right panel). Sudden gap drops at the critical strip widths $L_c\simeq 100$ and $L_c\simeq 220$nm yield maximum charge conductance regardless the value of $\mu_F$. This phenomenon does not occur for $E_g^\Gamma$.
Gap drops also occur when varying the Rashba term $\xi=15.6|e\mathcal{E}_z|$ by applying a perpendicular electric field $\mathcal{E}_z$, as shown in Fig. \ref{Fig:conductanceRashba}. For a strip width of $L=200$~nm, the gap drops down to $E_g\simeq 0.01$~meV for an electric field of $|\mathcal{E}_z|=22.4$~mV/nm (that is, $\xi\simeq 350$meV.nm), and the charge conductance rises to $G\simeq 0.9$. As suggested by \cite{Liu_PhysRevLett.100.236601,Cheng_2014}, if it is possible to have two independent control gates, one for the SIA and  other to change the Fermi energy level, then the variation of the charge conductance as function of the chemical potential ($\mu_F$) would be useful to design a QSH field effect transistor.

\section{Edge state localization properties} \label{secedgeloc}

We now proceed to analyze the localization properties of the four near-gap  states  $\{\Psi_{1c},\Psi_{1v},\Psi_{2c},\Psi_{2v}\}$, both in position $y$ and momentum $k=k_x$ independent spaces, each one of them  taking the form given in \eqref{4spinor}. 

\subsection{Analysis of near-gap state probability densities}

Let us firstly consider probability densities 
\begin{eqnarray}
|\Psi(k,y)|^2&=&|\psi_{\uparrow E}(k,y)|^2+|\psi_{\uparrow H}(k,y)|^2\nn\\ &+&|\psi_{\downarrow E}(k,y)|^2+ |\psi_{\downarrow H}(k,y)|^2.\label{normapsi}
\end{eqnarray}
 and normalize them according to $\int_0^L dy |\Psi(k,y)|^2=1$ (in the continuum limit $a\to 0$ of small lattice constant).

\begin{figure}
	\begin{center}
		\includegraphics[width=4cm]{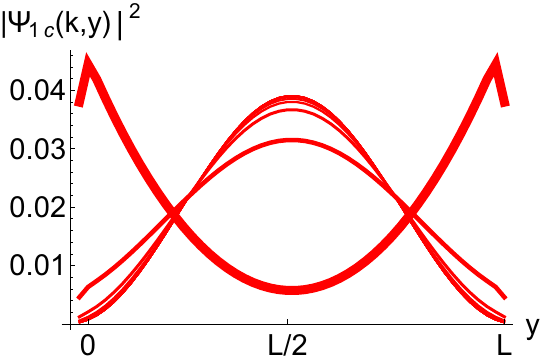}
		\includegraphics[width=4cm]{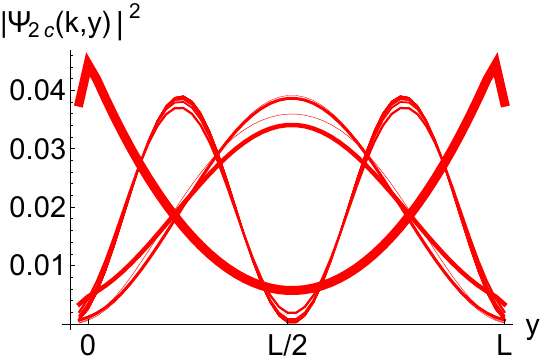}\\[2mm]
		\includegraphics[width=4cm]{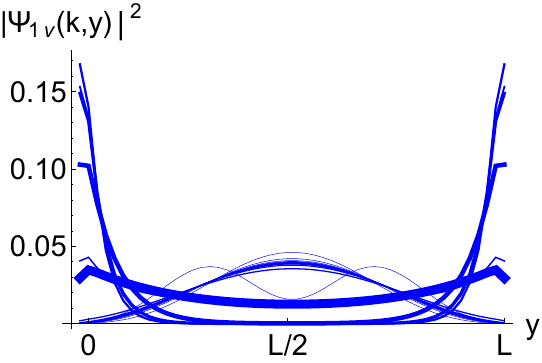}
		\includegraphics[width=4cm]{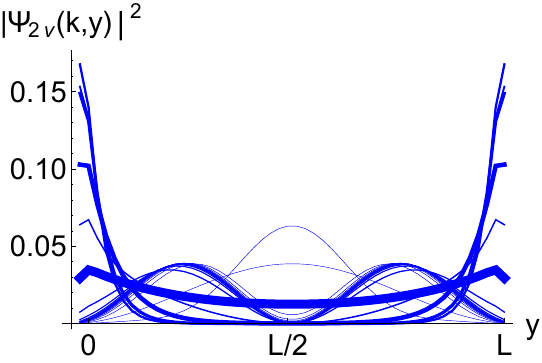}
	\end{center}
	\caption{Probability density $|\Psi_{c,v}(k,y)|^2$ of the four 4-spinor near-gap states $\{\Psi_{1c},\Psi_{1v},\Psi_{2c},\Psi_{2v}\}$ (conduction in red and valence in blue) as a function of  position $y\in [0,L], \,L=100$~nm, for several values of $k\in[0,\pi/a)$ (symmetric in $k$). The thickness of the probability density curves decreases with $k$, the thickest one corresponding to $k=0$. The maximum probability density  at the edges $y=0,L$ corresponds to:  $k\simeq\pm 0.21\,\mathrm{nm}^{-1}$ for valence band (blue) and $k=0$ for conduction band (red) states. See later in Figure \ref{Fig:3Density} for a 3D view. }
	\label{Fig:Density}
\end{figure}

In Fig. \ref{Fig:Density} we represent the probability densities  $|\Psi(k,y)|^2$ of the four near-gap states  $\{\Psi_{1c},\Psi_{1v},\Psi_{2c},\Psi_{2v}\}$ as a function of $y$ for several values of the momentum $k$ (varying curve thickness). They turn out to be symmetric in $k$, that is, $|\Psi_{c,v}(k,y)|^2=|\Psi_{c,v}(-k,y)|^2$, so that we take $k\in[0,\pi/a)$ for these plots. Maximum localization at the edges $y=0,L$ for valence states (in blue color) occurs at $k\simeq \pm 0.21\,\mathrm{nm}^{-1}$ (see also later in Fig. \ref{Fig:3Density} for a 3D plot), while for conduction states (in red color) it occurs at $k=0$ as an isolated value (see also later in Fig. \ref{Fig:3Density} for a 3D plot). For these particular values of $k$, valence band states  are more localized at the boundaries $y=0,L$ than conduction band states (in red color), approximately by a factor of four times, a fact which is also captured by the IPR later in Figure \ref{Fig:IPRk1vc} and 3D density plot in Figure \ref{Fig:3Density}.  

\begin{figure}[h]
	\begin{center}
		\includegraphics[width=\columnwidth]{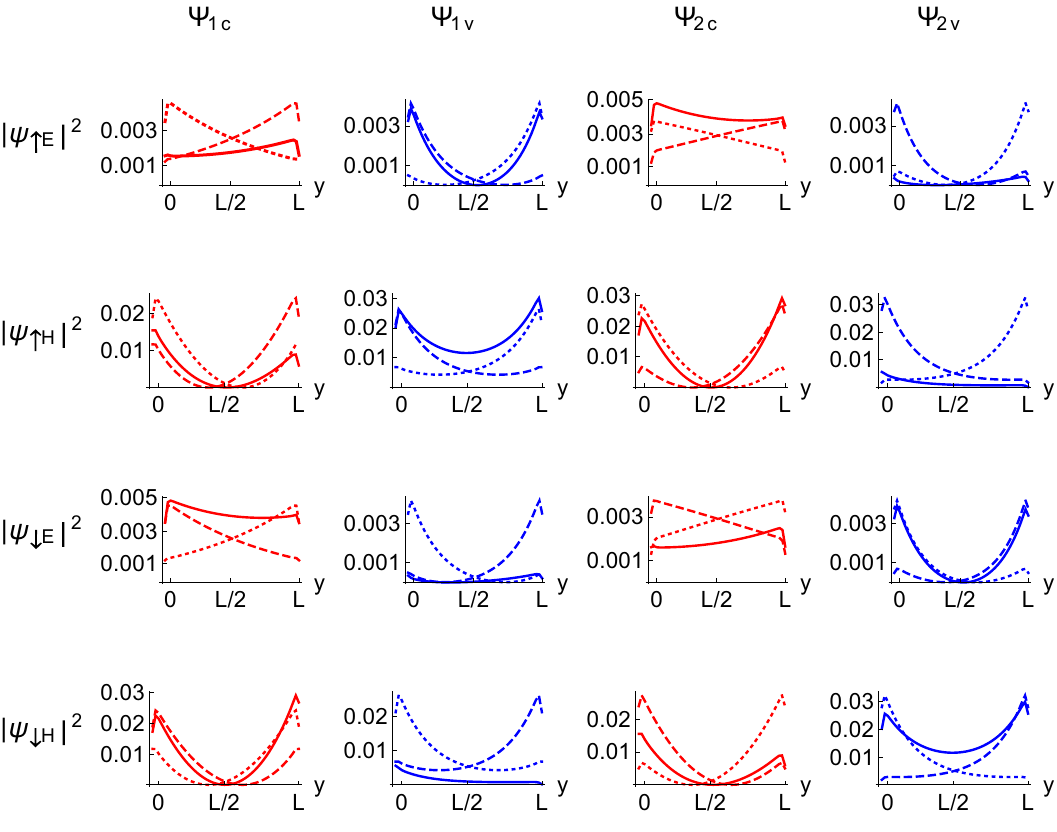}
	\end{center}
	\caption{Component-wise probability densities  of the four near-gap states $\{\Psi_{1c},\Psi_{1v},\Psi_{2c},\Psi_{2v}\}$, as a function of $y\in[0,L]$ ($L=100$nm), for three values of the momenta:  $k_x=- 0.005\,\mathrm{nm}^{-1}$ (dashed)  $k_x= 0.005\,\mathrm{nm}^{-1}$ (dotted) and $k_x=0\,\mathrm{nm}^{-1}$ (solid). Conduction in red and valence in blue colors.  The spin polarization structure of the QSH edge states can be glimpsed.}
	\label{Fig:DensityComponents}
\end{figure}

\begin{figure}
	\begin{center}
		\includegraphics[width=8cm]{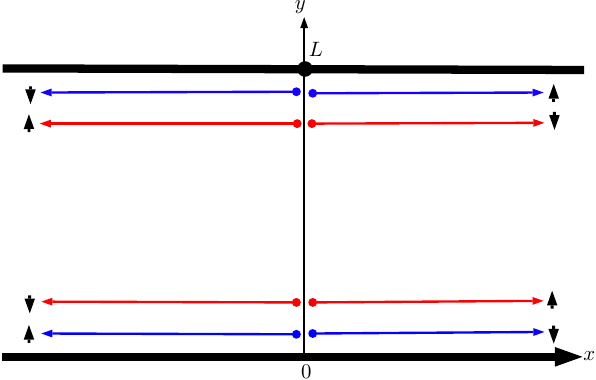}
	\end{center}
	\caption{Spin-polarization of edge states in a strip geometry of width $L$: the spin $s$ ($\uparrow=1, \downarrow=-1)$ is aligned with the propagation direction given by ${k_x}$ so that the helicity $h=\mathrm{sign}(s k_x)$ is $p=1$ for valence (blue) states at the edge $y=L$ and conduction (red) states at the edge $y=0$, and $p=-1$ for valence states at $y=0$ and conduction states at $y=L$.}
	\label{Fig:spinpol}
\end{figure}

For the time being, there is no clear momentum cutoff $k_c$ distinguishing between edge and bulk state regions.  Motivated by the analytic approach depicted in Fig. \ref{fig:comparison0}, we focus on low momentum  values like, for example, $k=0, \pm 0.005\,\mathrm{nm}^{-1}$,  where we hope to observe an edge state behavior. Indeed, 
a separated study of the four probability density components \eqref{4spinor} of the 4-spinor near-gap states $\{\Psi_{1c},\Psi_{1v},\Psi_{2c},\Psi_{2v}\}$ is shown in Fig. \ref{Fig:DensityComponents}. 
Note that, although $|\Psi_{c,v}(k,y)|^2$ does not depend on the sign of $k$, each component of $\Psi$ does. Spin down valence and spin up conduction component states are localized  at $y=L$ for $k<0$  and at $y=0$ for $k>0$, whereas spin up valence and spin down conduction component states are localized  at $y=0$ for $k>0$  and at $y=L$ for $k<0$. Therefore, there is a symmetry in $h=\mathrm{sign}(k s)$ (the helicity, with spin $s=\pm 1$). This is a  
reflect of the already commented spin polarization structure of the QSH 
edge states (remember Fig. \ref{probdensityfig}), experimentally observed in \cite{Brune2012} . See  also Fig. \ref{Fig:spinpol} for a pictorial representation of edge currents. Edge states are not chiral (since they propagate in both directions at both edges $y=0,L$), but they are ``spin filtered'' \cite{KaneMele05},  since electrons with opposite spins propagate in opposite directions.   For $k=0$, the probability density components  (solid curves in Fig. \ref{Fig:DensityComponents})  show a more balanced behavior at both edges $y=0,L$ in position space, as it was already noted in Fig.  \ref{probdensityfig} . 

Just looking at Hamiltonian spectra in Fig. \ref{Fig:Spectrum} or density distributions in Fig. \ref{Fig:Density}, it is difficult to establish a clear criterion for distinguishing between edge and  bulk states as $k$ moves inside the interval $[0,\pi/a]$. In Ref. \cite{KonigJPSJapan2008} a cutoff $k_c$,  bounding the edge state region  for the uncoupled tight-binding Hamiltonian,   is proposed, above which edge states are not longer normalizable and touch the bulk (extended) states.  A similar analysis for the SOC case becomes much more difficult and we have not found any additional result concerning this case in the literature.   Here we propose a cutoff for the coupled case  attending to a practical indicator of localization that has been utilized in a variety of settings, called the inverse participation ratio (IPR). This information measure is closely related to others like the Rényi entropy. Let us remind its definition.

\subsection{IPR and localization in position space $y$ for a given momentum $k_x$}

In general, the IPR  measures the spreading of the expansion of a normalized vector 
$|\Psi\ra=\sum_{n=1}^N c_n|n\ra$ in a given basis $\{|n\ra, n=1,\dots,N\}$. For a normalized vector $|\Psi\ra$, the probabilities  $p_n=|c_n|^2$ to find $|\Psi\ra$ in a given basis state $|n\ra$ fulfil $\sum_{n=1}^N p_n=1$.    The inverse participation ratio of $\Psi$ is defined as 
$$\mathrm{IPR}_\Psi=\sum_{n=1}^N p_n^2=\sum_{n=1}^N |c_n|^4.$$
It reduces to the square of the $\alpha$-norm $\|\Psi\|_\alpha=(\sum_{n=1}^N p_n^\alpha)^{1/\alpha}$ for $\alpha=2$. Moreover, it is also related to the R\'enyi entropy of order $\alpha$ 
$$ R^\alpha_\Psi=\frac{1}{1-\alpha}\log \sum_{n=1}^Np_n^\alpha,$$
for $\alpha=2$. R\'enyi entropy generalizes the quantum mechanical $\alpha\to 1$ von Neumann entropy. The smallest value  $\mathrm{IPR}_\psi=1/N$ corresponds to an equally weighted superposition $|c_n|=1/\sqrt{N}, \forall n$ (fully delocalized state), whereas the highest value  $\mathrm{IPR}_\psi=1$ corresponds to a perfectly localized state, that is, $c_n=\delta_{n,n_0}$ for some $n_0$.  

The concept can be similarly extended to the representation of the wavevectors in position, momentum or other spaces, so as to study the spreading, or the localization, of the vector in the given space basis. For example, for the case of a free particle in a box $y\in[0,L]$, the wave function $\psi_m(y)=\sqrt{2/L}\sin(m\pi y/L)$, normalized according to $\int_{0}^Ldy|\psi_m(y)|^2=1$,  has an  $\mathrm{IPR}_{m}=\int_{0}^Ldy|\psi_m(y)|^4=3/(2L)$.  For example, for a strip width of $L=100$~nm, we have $\mathrm{IPR}_m=0.015$.
 
 In our case, the IPR of  4-spinor $\Psi(k,y)$, normalized as $\int_0^L dy |\Psi(k,y)|^2=1$,  in  position $y$  space for each value of the momentum $k$ is given by

\begin{equation}
 \mathrm{IPR}_\Psi(k)=\int_0^L dy |\Psi(k,y)|^4, 
\end{equation}
where now we understand
\begin{equation}
 |\Psi|^4=|\psi_{\uparrow E}|^4+|\psi_{\uparrow H}|^4+|\psi_{\downarrow E}|^4+ |\psi_{\downarrow H}|^4.
\end{equation}

\begin{figure}
	\begin{center}
		\includegraphics[width=8.5cm]{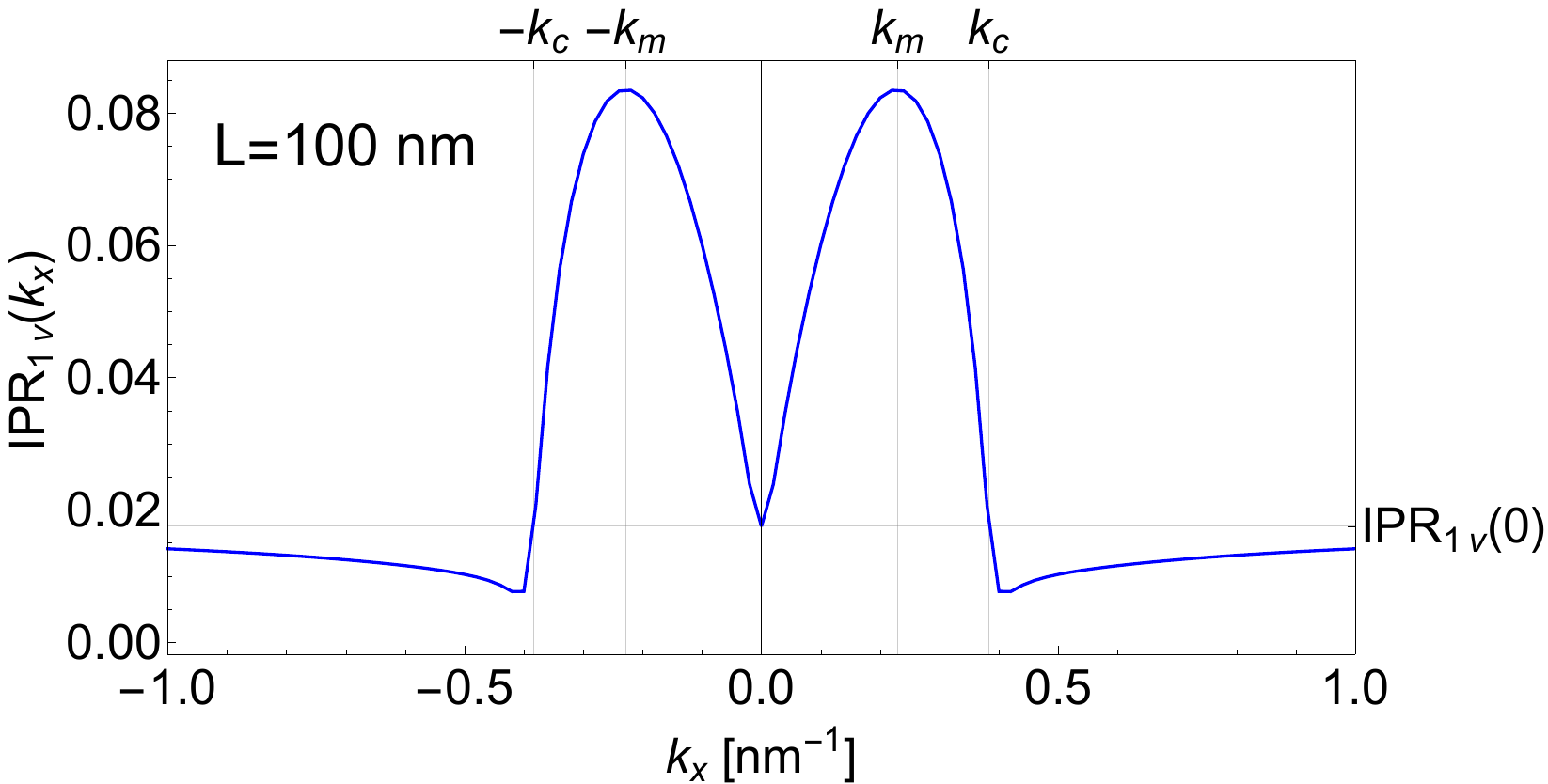}\\
		\includegraphics[width=8.5cm]{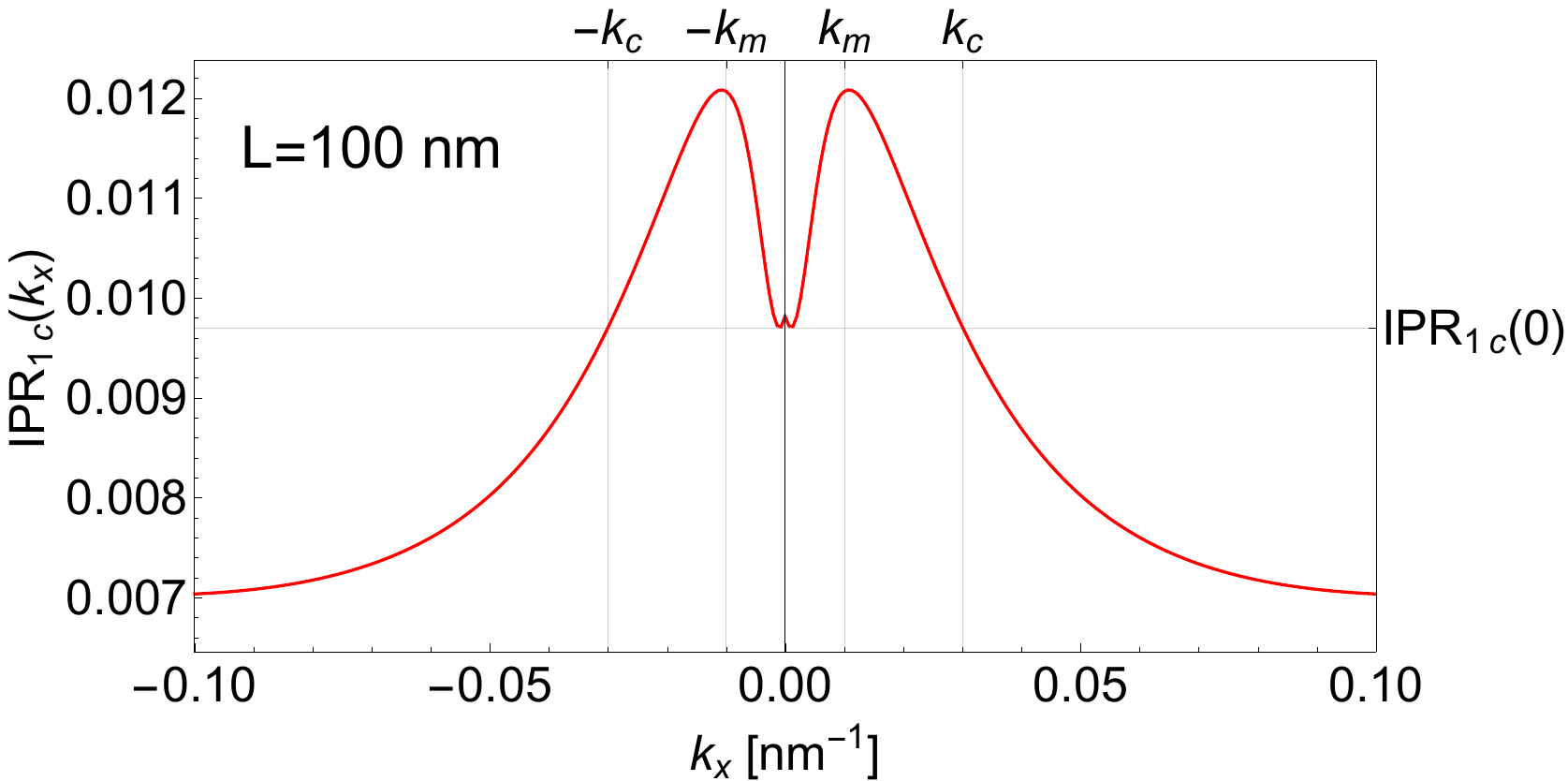}
	\end{center}
	\caption{Localization $\text{IPR}_{\Psi}(k)$ of the first near-gap valence $\Psi_{1v}$ (upper panel, blue)  and conduction $\Psi_{1v}$ (lower panel, red)  in position space $y$  as a function of the  momentum $k_x$  for a strip width of $L=100$~nm.  Maximum $\text{IPR}_{1v}(k)$ is attained at $\pm k_m\simeq \pm 0.229\,\mathrm{nm}^{-1}$ and maximum $\text{IPR}_{1c}(k)$ is attained at $\pm k_m\simeq \pm 0.01\,\mathrm{nm}^{-1}$. Cutoff momentum $k_c$ is defined as $\text{IPR}_{\Psi}(k_c)=\text{IPR}_{\Psi}(0^\pm)$, which gives $k_c\simeq 0.383\,\mathrm{nm}^{-1}$ for valence and $k_c\simeq 0.03\,\mathrm{nm}^{-1}$ for conduction first near-gap states.}
		\label{Fig:IPRk1vc}
\end{figure}

Fig. \ref{Fig:IPRk1vc} displays $\mathrm{IPR}_\Psi(k)$ for the  first near-gap valence  $\Psi_{1v}$ and conduction $\Psi_{1c}$ states  (similar results  for $\Psi_{2v}$ and  $\Psi_{2c}$) as a function of $k_x$ for $L=100$~nm. Its shape is symmetric, as expected. Note that localization of valence and conduction states occurs here at different scales (roughly a factor of ten). 
 $\Psi_{1v}$ shows maximum localization (maximum $\text{IPR}$)  at $\pm k_m\simeq \pm 0.229\,\mathrm{nm}^{-1}$ and $\Psi_{1c}$  at $\pm k_m\simeq \pm 0.01\,\mathrm{nm}^{-1}$. 
 
 Remember that  a cutoff $k_c$ was defined in Fig. \ref{fig:comparison0} and Eq. \eqref{kcutoffL} for the analytic approach as the value of $k_x$ at which edge and bulk states merge. That is, after the replacement $k_y\to-\mathrm{i} \lambda$ done in the Hamiltonian $H(k_x,k_y)$ at the beginning of  Section \ref{analyticedge} to look for edge localized solutions $\Psi(y)\sim e^{\lambda y}$ of the Schr\"odinger equation, the criterion for edge-bulk transmutation was $\lambda=0=\mathrm{i} k_y$, providing  the cut-off values of $k_x$ given in equation \eqref{kcutoffL}.
 
 In the tight-binding approach, our proposal for a proper definition of $k_c$ for a given near-gap state $\Psi$ is when $\text{IPR}_{\Psi}(k_c)$ is comparable to $\text{IPR}_{\Psi}(0^\pm)$, that is
 \be 
 \text{IPR}_{\Psi}(k_c)=\lim_{k\to 0^\pm}\text{IPR}_{\Psi}(k).\label{cutoffdef}
 \ee
This definition requires in general to take the one-sided limit  $k\to 0^\pm$ to avoid those cases in which there is a removable discontinuity of $\text{IPR}_{\Psi}(k)$ at $k=0$. Our proposal for $k_c$ is physically justified by the fact that low momentum $k\to 0$ states correspond to \emph{translationally  invariant} (highly delocalized, low IPR) configurations. Therefore, for  near-gap (low momentum) states $\Psi$,  a reasonable definition of $k_c$ is when $\text{IPR}_{\Psi}(k_c)$ attains low values comparable to $\text{IPR}_{\Psi}(0^\pm)$.

The condition  \eqref{cutoffdef} gives $k_c\simeq 0.383\,\mathrm{nm}^{-1}$ for valence and $k_c\simeq 0.03\,\mathrm{nm}^{-1}$ for conduction first near-gap states for $L=100$~nm, as depicted in Fig. \ref{Fig:IPRk1vc}. 
The value of the cutoff $k_c$ depends on $L$. In Table  \ref{tabla2} we give several values of $k_c$ for different strip widths $L$. The value of $k_c$ for valence states seems to have a lower dependence on $L$ than for conduction states.

Note that these cutoff $k_c$ values differ from those obtained by formula \eqref{kcutoffL} which, for example, give $k_c\simeq 0.004\,\mathrm{nm}^{-1}$ for conduction first near-gap states for $L=100$~nm. This is because, in addition to the fact that we are using a different (numeric, tight-binding) approach, we are also considering full SOC (BIA plus SIA) interaction, which was absent in the analytic approach of Section \ref{analyticedge}.

\begin{table}[]
\begin{center}
\begin{tabular}{l|c|c|c|}
              $k_c$ {[}nm$^{-1}${]}              & \multicolumn{1}{l|}{$L=100$~nm} & \multicolumn{1}{l|}{$L=200$~nm} &  \multicolumn{1}{l|}{$L=250$~nm} \\ \hline
\multicolumn{1}{|l|}{$\Psi_{1c}$} & $0.03$                                 & $0.011$              &      $0.007$              \\ \hline
\multicolumn{1}{|l|}{$\Psi_{1v}$} & $0.383$                                 & $0.388$                &     $0.389$             \\ \hline 
\end{tabular}
 \end{center}
\caption{Edge state momentum cutoff $k_c$  for  $\Psi_{1c}$ and $\Psi_{1v}$ as a function of the  strip  width for $L=100, 200$ and $250$~nm.}
\label{tabla2}
\end{table}

\begin{figure}
	\begin{center}
		\includegraphics[width=4.2cm]{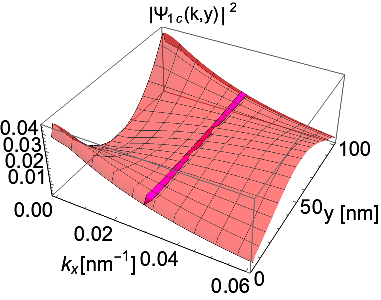}
		\includegraphics[width=4.2cm]{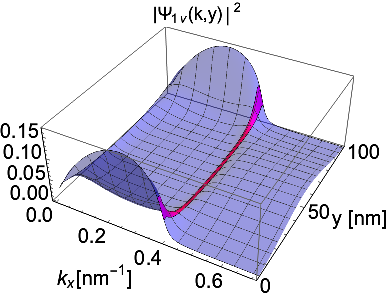}
	\end{center}
	\caption{3D version of the probability density  in Figure \ref{Fig:Density} for first conduction $|\Psi_{1c}|^2$ (red) and valence $|\Psi_{1v}|^2$ (blue) near-gap states  as a function of $k_x$ and $y\in[0,L]$ for $L=100$~nm. The cutoff momentum $k_c$ in Table \ref{tabla2} distinguishing edge from bulk regions is highlighted in magenta.}	
	\label{Fig:3Density}
\end{figure}

Fig. \ref{Fig:3Density} shows a 3D version of the probability density $|\Psi(k,y)|^2$ in Figure \ref{Fig:Density}. Only  positive momentum values are represented, since the probability density is symmetric in $k_x$. The proposed cutoff $k_c$ separating edge from bulk regions is highlighted in magenta color.

The probability densities $|\Psi_{1c,1v}(k_c,y)|^2$, in magenta color in Fig. \ref{Fig:3Density}, correspond to the flattest ones in the real $y$-space, that is, the quantity 
\begin{equation}
 \textrm{Max}_{y\in[0,L]}|\Psi(k,y)|^2-\textrm{Min}_{y\in[0,L]}|\Psi(k,y)|^2
\end{equation}
exhibits a minimum value for $k\simeq k_c$. Note that this criterium approximately coincides with that of Eq. \eqref{cutoffdef}.

 \subsection{IPR and localization in momentum space $k_x$ for a given position $y$}

\begin{figure}
	\begin{center}
		\includegraphics[width=4.2cm]{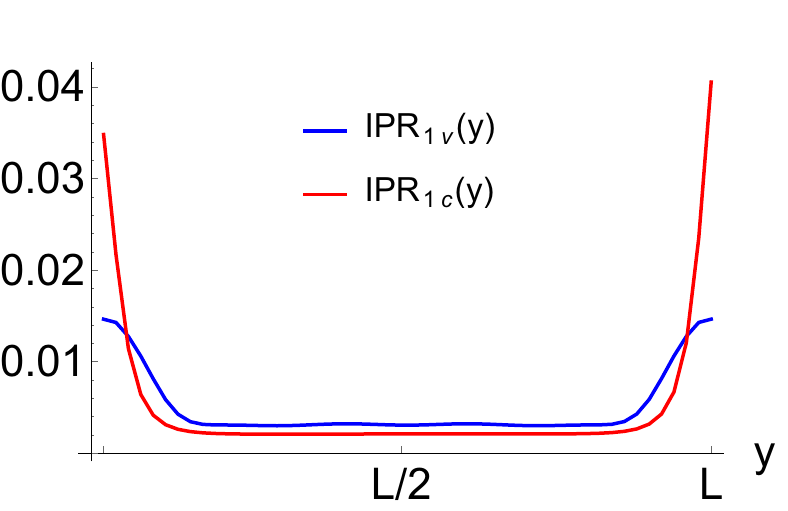}
		\includegraphics[width=4.2cm]{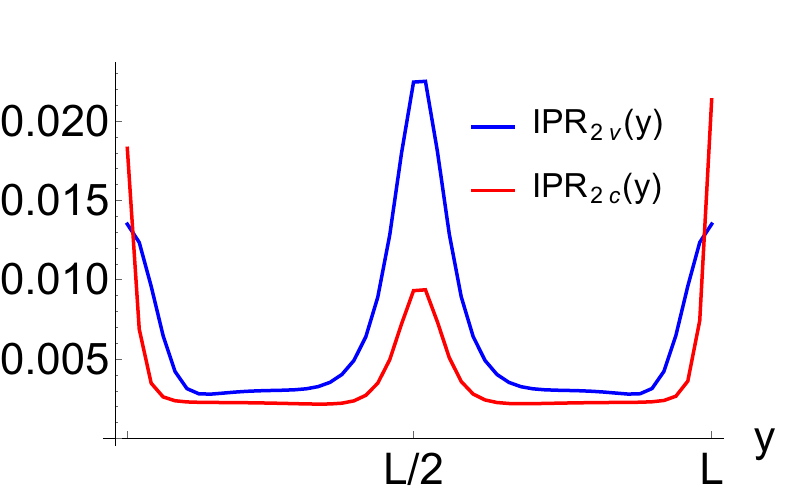}
	\end{center}
	\caption{Localization of conduction (red) and valence (blue) near-gap states in momentum space for each value of the position ($y$) as measured by the inverse participation ratio for a strip width of $L=100$~nm. Near-gap states participate of less momenta at $y=0,L$. $\Psi_{2c}$ and  $\Psi_{2v}$ also show higher IPR (momentum localization) at the center of the strip $y=L/2$.
	}
	\label{Fig:IPRy}
\end{figure}

For completeness, we also analyze the spreading of the expansion of near-gap states in momentum space $k$ for a given position $y$. To do that, we have to normalize 4-spinors as $\int_{-\pi/a}^{\pi/a} dk |\Psi(k,y)|^2=1$ and define 
\begin{equation}
 \mathrm{IPR}_\Psi(y)=\int_{-\pi/a}^{\pi/a} dk |\Psi(k,y)|^4.
\end{equation}
Fig. \ref{Fig:IPRy} shows that near-gap states participate of less and less momenta $k$ as we approach the boundaries $y=0,L$ (higher IPR), since momentum is localized around $\pm k_m\simeq \pm 0.229\,\mathrm{nm}^{-1}$, as it was noted before in Fig \ref{Fig:IPRk1vc}. The second near-gap conduction $\Psi_{2c}$ and valence $\Psi_{2v}$ states  also participate of less and less momenta $k$  (higher IPR) towards  the center of the strip $y=L/2$. 

The IPR concept is related to the purity of a density matrix, which measures the degree of entanglement of a given physical state. In the next section we study entanglement properties of our edge states.

\section{Spin probabilities and spin-band entanglement measures}\label{secentang} 

In order to compute quantum correlations in our system, we shall use two different entanglement measures. 

\subsection{Reduced density matrix, spin probabilities and linear entropy}

Let $\rho=|\Psi\ra\la\Psi|$ the $4\times 4$ density matrix $\rho$ corresponding to a normalized 4-spinor state \eqref{4spinor}. Denoting the 4-spinor $\Psi(k,y)$ column 4-vector as a function of position $y$ and momentum $k$, the $4\times 4$ density matrix at $(k,y)$ acquires the form
\begin{equation}
 \rho(k,y)=\frac{\Psi(k,y)\Psi^\dag(k,y)}{\Psi^\dag(k,y)\Psi(k,y)},
\end{equation}
where we are normalizing by the scalar quantity $\Psi^\dag(k,y)\Psi(k,y)=|\Psi(k,y)|^2$ in \eqref{normapsi} in order to have $\tr(\rho(k,y))=1$ at each point $(k,y)$. 
The 16 density matrix entries $\rho_{ij}, i,j=1,2,3,4$ are referenced to the basis
\begin{eqnarray}
 |1\ra&=&|\uparrow\ra\otimes|E\ra, \quad |2\ra=|\uparrow\ra\otimes|H\ra, \nn\\ |3\ra&=&|\downarrow\ra\otimes|E\ra,\quad  |4\ra=|\downarrow\ra\otimes|H\ra.
\end{eqnarray}
The $2\times 2$ reduced density matrix (RDM) valued  $\varrho(k,y)$ to the spin subsystem is obtained by tracing out the orbital $E, H$ degrees of freedom
\be
\varrho=\tr_{EH}(\rho)=\left(
\begin{array}{cc}
 \rho_{11}+\rho_{22} & \rho_{13}+\rho_{24} \\
 \rho_{31}+\rho_{42} & \rho_{33}+\rho_{44} \\
\end{array}
\right).
\ee
The diagonal components of the RDM 
\begin{equation}
 \varrho_{11}=\rho_{11}+\rho_{22}=P_\Psi(\uparrow), \quad \varrho_{22}=\rho_{33}+\rho_{44}=P_\Psi(\downarrow),
\end{equation}
represent the probabilities of finding the electron with spin up or down at each point $(k,y)$,   respectively, whereas the modulus of the off-diagonal elements 
\begin{equation}
 |\varrho_{12}|=|\varrho_{21}|=|\rho_{13}+\rho_{24}|=P_\Psi(\uparrow\to\downarrow),
\end{equation}
represent the spin-transfer probability amplitudes (also called coherences in quantum information jargon).

\begin{figure}
	\begin{center}
		\includegraphics[width=4cm]{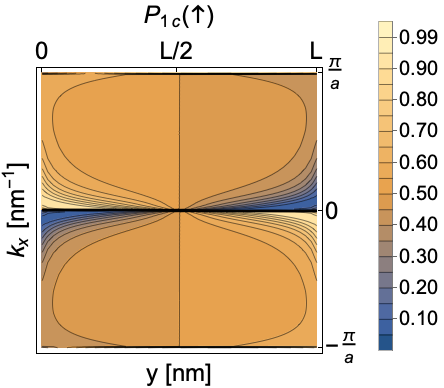}\quad\includegraphics[width=4cm]{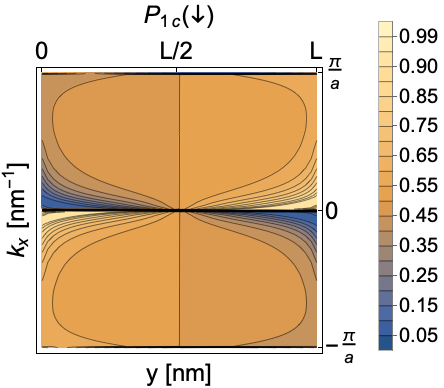}\\[2mm]
		\includegraphics[width=4cm]{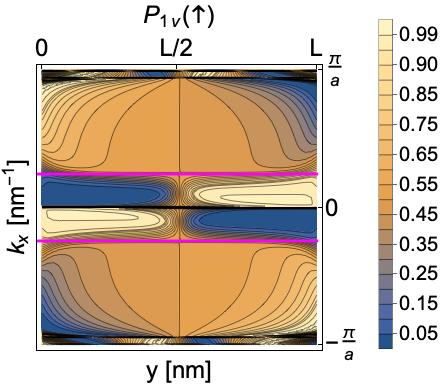}\quad\includegraphics[width=4cm]{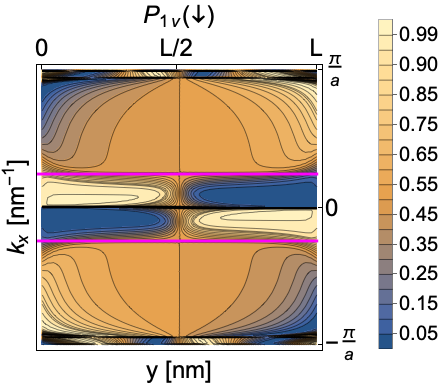}
	\end{center}
	\caption{Probabilities to find the electron with spin $\uparrow$ and down $\downarrow$ for the first  conduction $\Psi_{1c}$ and valence $\Psi_{1v}$ near-gap  states as a function of momentum in the entire Brillouin zone, $k\in(-\pi/a,\pi/a)$, and position $y\in(0,L)$ for a strip width of $L=100$~nm. Lighter zones correspond to higher probability. The cutoff momentum $k_c=0.383\mathrm{nm}^{-1}$ in Table \ref{tabla2} distinguishing edge from bulk regions is highlighted in magenta.} 
	\label{Fig:Probabilityspin_edge_allBZ}
\end{figure}

In Fig. \ref{Fig:Probabilityspin_edge_allBZ} we plot the probabilities $P_\Psi(\uparrow)$ and $P_\Psi(\downarrow)$ for the first conduction $\Psi_{1c}$ and valence $\Psi_{1v}$ near-gap states as  a function of $(k,y)$ in the entire Brillouin zone $[-\pi/a,\pi/a]$. Lighter colors represent higher probability zones.  Probability densities are unbalanced at the boundaries $y=0,L$ for low momenta $k$, depending on the propagation direction given by the sign of $k$. This is again a reflection of the existence of counterpropagating modes of opposite spin at the edges (helical edge modes), which is consistent with the experimental confirmation in \cite{Brune2012} that the transport in the edge channels is spin polarized.  This is so because, at  each edge of the strip, the sign $\text{sign}(ks)$ is well defined, being $+1$ for the  $y=0$ edge and $-1$ for the $y=L$ edge.  
\begin{figure}
	\begin{center}
		\includegraphics[width=4.1cm]{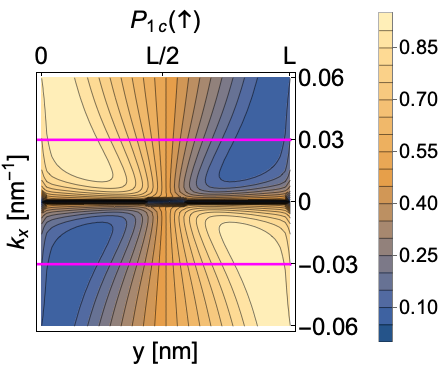}\quad\includegraphics[width=4.1cm]{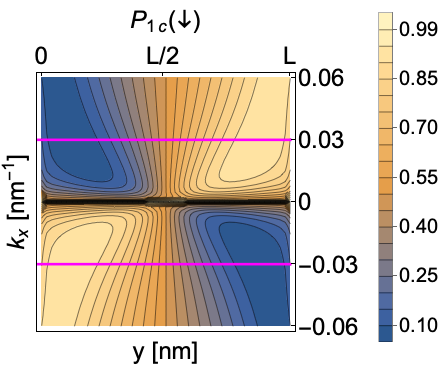}
	\end{center}
	\caption{A zoom of conduction probabilities  $P_{1c}(\uparrow) $ and $P_{1c}(\downarrow)$  of Figure  \ref{Fig:Probabilityspin_edge_allBZ}  in the edge state region neighborhood  $k\in[-2k_c,2k_c]$ for the cutoff $k_c\simeq 0.03,\mathrm{nm}^{-1}$ (in magenta)  in Table \ref{tabla2}.} 
	\label{Fig:Probabilityspin_edge}
\end{figure}
The magenta line in Fig. \ref{Fig:Probabilityspin_edge_allBZ} (lower panel) corresponds to the  cutoff momentum $k_c=0.383\mathrm{nm}^{-1}$ in Table \ref{tabla2} (valence states) distinguishing edge from bulk regions. For conduction states we provide a zoom of the edge state region in Fig. \ref{Fig:Probabilityspin_edge} since, unlike for valence states, the cutoff $k_c=0.03\mathrm{nm}^{-1}$ is too small (compared to $\pi/a$) to be appreciated. Fig. \ref{Fig:Probabilityspin_edge_allBZ}  says that the spin polarization structure, as evidenced by the diagonal components $P_\Psi(\uparrow)$ and $P_\Psi(\downarrow)$ of the RDM $\varrho$, is characteristic of the edge-state region $|k|\lesssim k_c$, blurring upon entering the bulk-state region $|k|\gtrsim k_c$. This could be another alternative criterion to distinguish edge-state from bulk-state regions.

\begin{figure}
	\begin{center}
		\includegraphics[width=4.2cm]{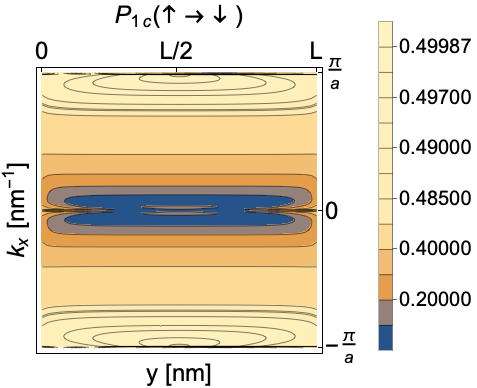} 
		\includegraphics[width=4.2cm]{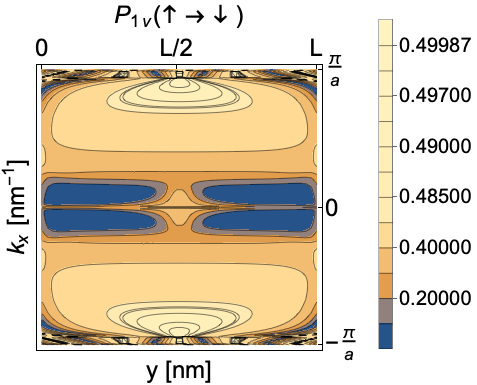}
	\end{center}
	\caption{Spin-transfer probability amplitudes for first  conduction $\Psi_{1c}$ and valence   $\Psi_{1v}$ edge states as a function of momentum $k\in(-\pi/a,\pi/a)$ and position $y\in(0,L)$ for a strip width of $L=100$~nm. Lighter zones correspond to higher transfer probability amplitudes. 
	}
	\label{Fig:Probabilitytransitionspin}
\end{figure}

In Fig. \ref{Fig:Probabilitytransitionspin} we plot spin-transfer probability amplitudes $P_\Psi(\uparrow\to\downarrow)$  for  $\Psi_{1c}$ and  $\Psi_{1v}$  as a function of $k$ and $y$ for a strip width of $L=100$~nm. The maximum probability $P_{1c}^{\mathrm{max.}}(\uparrow\to\downarrow)\simeq 1/2$ is attained at the center of the strip $y=L/2$ for $k\simeq \pm 1.54\,\mathrm{nm}^{-1}$, and the minimum probability $P_{1c}^{\mathrm{min.}}(\uparrow\to\downarrow)=0.003$ is attained at   $y=34$ and $y=66$~nm for $k\simeq \pm 0.1\,\mathrm{nm}^{-1}$. These extrema are quite flat, as can be perceived in Fig.  \ref{Fig:Probabilitytransitionspin}. Analogously, for the first valence edge state, there is a quite flat zone of maximum probability $P_{1v}^{\mathrm{max.}}(\uparrow\to\downarrow)\simeq 1/2$ around   $y=L/2$ and  $k\simeq \pm 1.4\,\mathrm{nm}^{-1}$, and of minimum probability $P_{1v}^{\mathrm{min.}}(\uparrow\to\downarrow)=0.02$ around   $y=22$ and $y=78$~nm for $k\simeq \pm 0.1\,\mathrm{nm}^{-1}$. 

The main conclusion that can be drawn from Fig. \ref{Fig:Probabilitytransitionspin}  is that spin-transfer probability amplitudes in the edge-state region $|k|\lesssim k_c$ (excluding perhaps a neighborhood of the $\Gamma$ point $k=0$) are much lower than in the bulk state region $|k|\gtrsim k_c$. Therefore, one could say that edge currents conserve spin better than bulk currents.

We now analyze the spin-band quantum correlations by means of the linear entropy, which is defined through the purity $\tr(\varrho^2)$ as
\be
S=1-\tr(\varrho^2).
\ee
Maximum entanglement means $S_\mathrm{max}=1/2$ for a $2\times 2$ RDM $\varrho$, whereas pure states have $S=0$.

\begin{figure}
	\begin{center}
		\includegraphics[width=4.3cm]{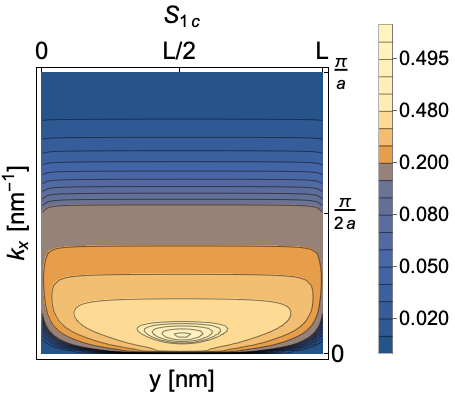}\includegraphics[width=4.3cm]{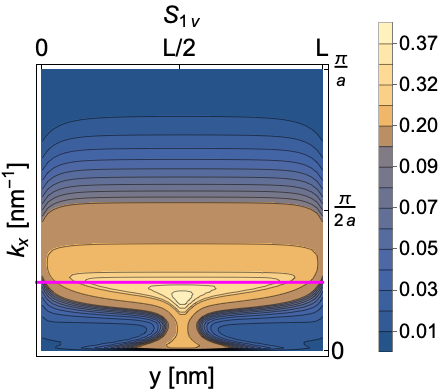}\\[2mm]
		\includegraphics[width=4.3cm]{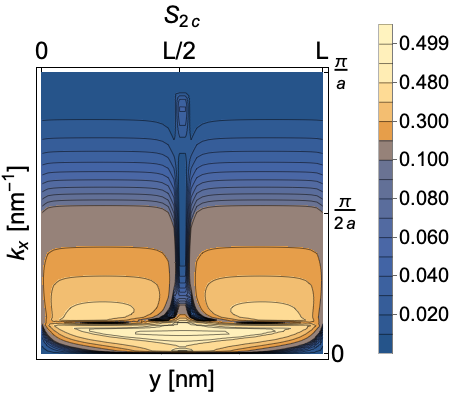}\includegraphics[width=4.3cm]{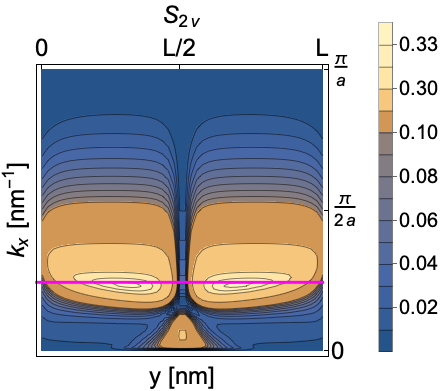}
	\end{center}
	\caption{Linear entropy $S$ for spin-band correlations of the four near-gap states $\Psi_i, i\in\{1c,1v,2c,2v\}$ as a function of momentum $k\in(0,\pi/a)$ ($S$ is symmetric in $k$ and $y$) and position $y\in(0,L)$ for a strip width of $L=100$~nm. Lighter zones correspond to higher entropy. The cutoff momentum $k\simeq 0.383\,\mathrm{nm}^{-1}$ in Table \ref{tabla2} distinguishing edge from bulk regions is highlighted in magenta color for valence states. }
	\label{Fig:Entanglement_allBZ}
\end{figure}

In Fig.  \ref{Fig:Entanglement_allBZ}  we show the linear entropies $S_i(k,y)$ of the four near-gap states $\Psi_i$, with $i\in\{1c,1v,2c,2v\}$, as a function of $(k,y)$ for a strip width of $L=100$~nm. The entropy is symmetric in $k$ and $y$, and we shall only show half of the interval in momentum space (that is $k\in(0,\pi/a)$). For $\Psi_{1c}$ and $\Psi_{2c}$, the maximum entanglement $S\simeq 1/2$ occurs at $y=L/2$ and $k\simeq 0.11\,\mathrm{nm}^{-1}$. For $\Psi_{1v}$, the maximum entanglement $S\simeq 0.38$ occurs at $y=L/2$ and $k\simeq 0.3\,\mathrm{nm}^{-1}$. For $\Psi_{1v}$, the maximum entanglement $S\simeq 0.38$ occurs at $y=L/2$ and $k\simeq 0.3\,\mathrm{nm}^{-1}$ whereas for $\Psi_{2v}$, the maximum entanglement $S\simeq 0.33$ occurs at $y\simeq 32$ and $y\simeq 68$~nm and $k\simeq 0.36\,\mathrm{nm}^{-1}$. In any case, entanglement shows a higher variability in the low momentum $k$ region below the cutoff $k_c$, where edge states lay, than in the bulk state region (with a quite low entropy uniformity). 

\begin{figure}
	\begin{center}
		\includegraphics[width=4.3cm]{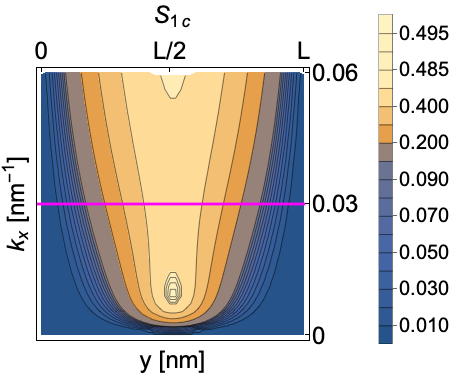}\includegraphics[width=4.3cm]{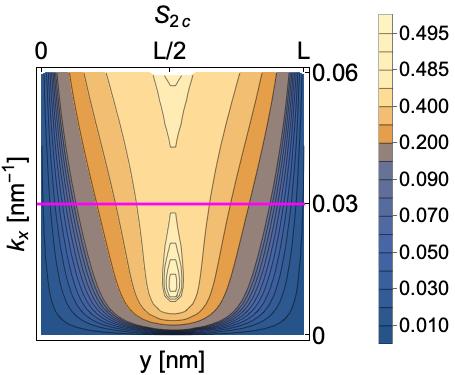}
	\end{center}
	\caption{A zoom of entropies $S_{1c} $ and $S_{2c}$  of Figure \ref{Fig:Entanglement_allBZ}  in the low momentum region  $k\in[0,2k_c]$ around  the cutoff $k_c\simeq 0.03,\mathrm{nm}^{-1}$ (in magenta)  in Table \ref{tabla2}
	}
	\label{Fig:Entanglementzoom}
\end{figure}

Fig.  \ref{Fig:Entanglementzoom} shows a zoom of the edge-state region for conduction states.

\subsection{Schlienz \& Mahler entanglement measure}\label{Schlienzsec}

We shall also briefly discuss other related  entanglement measure in the field of quantum information,  like the one proposed by Schlienz \& Mahler \cite{Mahler} related to a bipartite system of an arbitrary number $D$ levels (``quDits''). In our case, $D=2$ and a  qubit-qubit system will make reference to spin up-down and band E-H sectors. The entanglement measure is defined as follows. The $4\times 4$ density matrix $\rho$ is now written in terms of the 16 generators of the unitary group U(4), which can be written as tensor products of Pauli matrices like in \eqref{ham0} and \eqref{ham01}. More precisely
\begin{eqnarray}
\rho &=& \frac{1}{4}\sigma_0 \otimes \sigma_0+ \frac{1}{4} \sum_{k=1}^3 (\lambda^{(1)}_k \sigma_k \otimes \sigma_0+   \lambda^{(2)}_k \sigma_0 \otimes \sigma_k)  \nn\\ &&+  \frac{1}{4} \sum_{k, j} C^{(1,2)}_{kj} \sigma_k \otimes \sigma_{j} \, ,
\end{eqnarray}
with 
 \begin{eqnarray}
&\bm{\lambda}^{(1)} = \tr(\rho\,\bm{\sigma} \otimes \sigma_0) , \, \bm{\lambda}^{(2)} = \tr(\rho\,\sigma_0\otimes \bm{\sigma} ), &\nn\\ 
&{C}^{(1,2)} _{kj}= \tr(\rho\,{\sigma}_k \otimes {\sigma}_j).&
\end{eqnarray}
The vectors $\bm{\lambda}^{(1)}$ and $\bm{\lambda}^{(2)}$  denote the Bloch coherence vectors of the first qubit (spin  up-down) and the second qubit (band E-H) and the  $3 \times 3$ matrix ${C}^{(1,2)} $ accounts for qubit-qubit (spin-band) correlations. The RDM on the spin sector is 
\begin{equation}
\rho^{(1)}=\tr_2(\rho)=\frac{1}{2}\sigma_0+\frac{1}{2}\sum_{k=1}^3\lambda^{(1)}_k\sigma_k \otimes \sigma_0,
\end{equation}
and analogously on the band sector $\rho^{(2)}$. Comparing $\rho$ with the direct product $\rho^{(1)}\otimes \rho^{(2)}$, the difference comes from a $3\times 3$ entanglement matrix $M$ with components
\begin{equation}
{M}_{jk} = {C}^{(1,2)}_{jk} - \lambda^{(1)}_j \lambda^{(2)}_k ,\; j,k = 1,2,3.\label{MMahler}
\end{equation}
Based on $M$, Ref. \cite{Mahler} introduces a measure of ``qubit-qubit'' (spin-band)  entanglement given by the parameter 
\begin{equation}
B_\Psi= \frac{1}{3} \tr({M}^T {M} ).
\end{equation}
The parameter $B$ is bounded by $0\leq B \leq 1$ and carries  information about spin up and down correlations. The results for  $B$ provide an equivalent behavior to the linear entropy in Figure \ref{Fig:Entanglement_allBZ}, except for a scaling factor.

The conclusion here is that the behavior of quantum correlations in HgTe QWs appears to be independent of the specific entanglement measure employed.

\section{Correlation functions for real-space partitions}\label{ypartsec}

\begin{figure}
	\begin{center}
		\includegraphics[width=8.5cm]{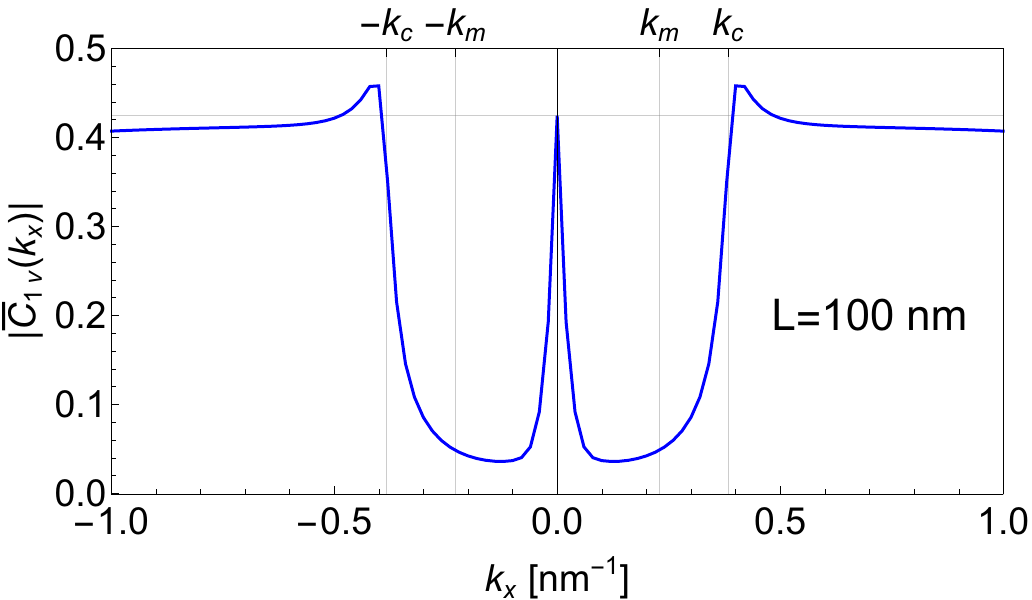}\\
		\includegraphics[width=8.5cm]{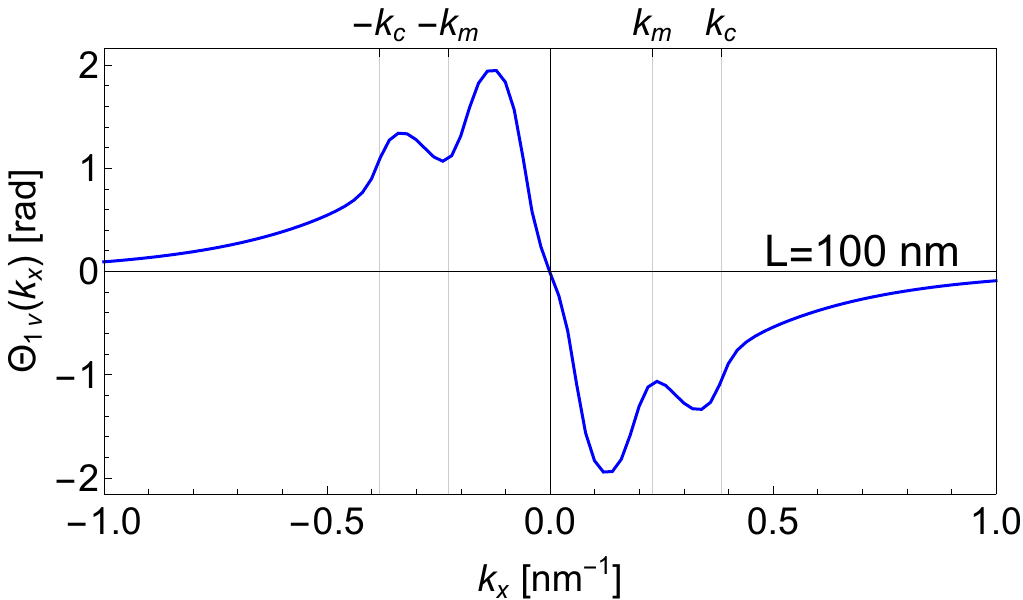}
	\end{center}
	\caption{Correlation fucntion  $C_{\Psi}(k)$ (modulus $|C_{\Psi}(k)|$ and argument $\Psi_{\Psi}(k)$)  of the  first near-gap valence state $\Psi_{1v}$ for a real-space partition $y\in [0,L/2]\cup [L/2,L]$ as a function of the momentum $k$. Maximum $k_m$ and cut-off $k_c$ momenta of Fig. \ref{Fig:IPRk1vc} (upper panel) are marked with vertical grid lines to match IPR with correlation functions. }
		\label{Fig:Ck1v}
\end{figure}

\begin{figure}
	\begin{center}
		\includegraphics[width=8.5cm]{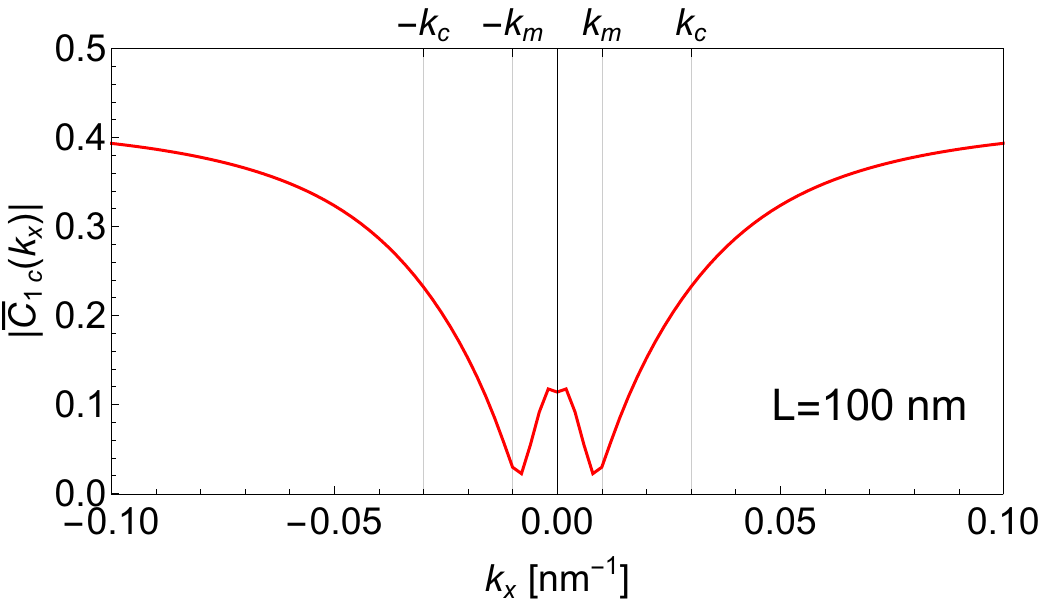}\\
		\includegraphics[width=8.5cm]{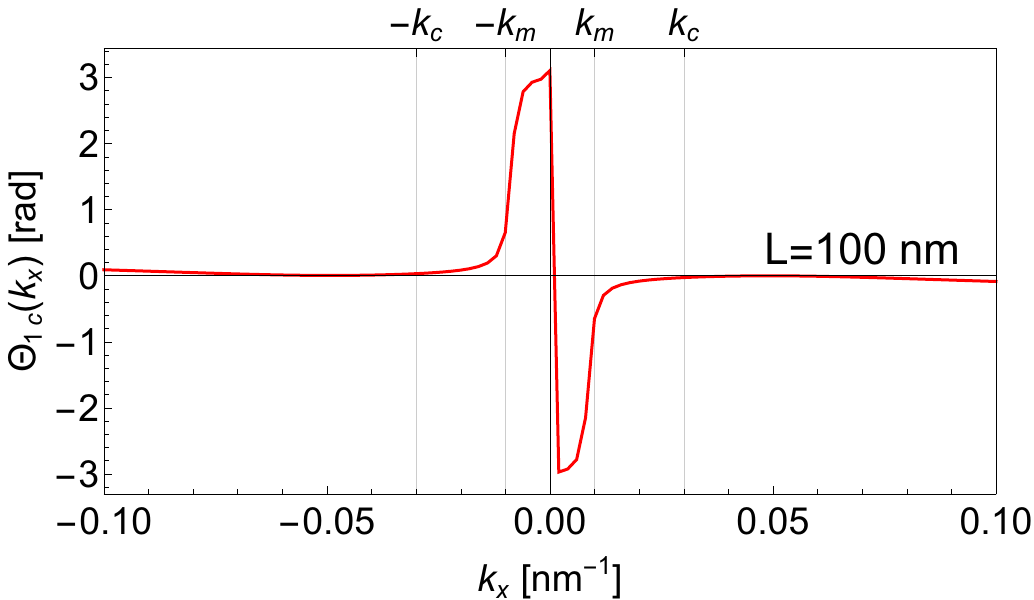}
	\end{center}
	\caption{Correlation fucntion  $C_{\Psi}(k)$ (modulus $|C_{\Psi}(k)|$ and argument $\Psi_{\Psi}(k)$)  of the  first near-gap conduction state $\Psi_{1c}$ for a real-space partition $y\in [0,L/2]\cup [L/2,L]$ as a function of the momentum $k$. Maximum $k_m$ and cut-off $k_c$ momenta of Fig. \ref{Fig:IPRk1vc} (lower panel) are marked with vertical grid lines to match IPR with correlation functions.  }
		\label{Fig:Ck1c}
\end{figure}

In this Section we propose a real-space partition $y\in [0,L_1]\cup [L_1,L]$ (that is, a bipartition of the system along the $y$-direction) to probe edge versus bulk behavior. Other real-space partitions appear in the literature \cite{PhysRevLett.96.110404,IngoPeschel_2003}. Let $|\Psi\ra$ a $4N$ component spinor state; for example, an eigenstate  of the tight-binding Hamiltonian \eqref{hamTB} for a strip width $L=aN$ and momentum $k=k_x$. The two-point correlation function for this state is
\begin{equation}
 C_{\Psi}(k,m,n)=\la\Psi| c^\dag_{k,m} c_{k,n}|\Psi\ra,
\end{equation}
where $c_{k,n}$ and $c^\dag_{k,n}$ are the annihilation and creation operators with momentum $k$ at lattice site $n$ which define the tight-binding Hamiltonian \eqref{hamTB}. For the real-space partition $y\in [0,L_1]\cup [L_1,L]$, the space correlation function is then
\begin{equation}
\overline{C}_\Psi(k)=\frac{\sum_{m=0}^{N_1}\sum_{n=N_1+1}^N C_{\Psi}(k,m,n)}{\sqrt{(N_1(N-N_1)}},
\end{equation}
with $N_1=L_1/a$. In the continuum limit (small lattice constant) it could be written as 
\begin{equation}
 \overline{C}_\Psi(k)=\frac{\int_0^{L_1}\int_{L_1}^L dy dy' \bar{\Psi}(k,y)\Psi(k,y')}{\sqrt{(L_1(L-L_1)}},
\end{equation}
for a normalized probability density \eqref{normapsi}. 
 Figures \ref{Fig:Ck1v} and \ref{Fig:Ck1c} show plots of the  correlation function (modulus and argument) 
 \begin{equation}
 \overline{C}_\Psi(k)=|\overline{C}_\Psi(k)| e^{\ic \Theta_\Psi(k)}
 \end{equation}
 for the space-partition $L_1=L/2$   and the  first near-gap valence $\Psi_{1v}$ (blue)  and conduction $\Psi_{1c}$ (red) states,    as a function of the  momentum $k=k_x$  for a strip width of $L=100$~nm. Comparing with Fig.  \ref{Fig:IPRk1vc}, we see that higher IPR values correspond to lower space-correlation modulus $|\overline{C}_\Psi(k)|$ values. Moreover, space-correlation arguments/angles $\Theta_\Psi(k)$ attain their maximum absolute values in the edge-state (high IPR) region,  exhibiting positive (resp. negative) values for negative (resp. positive) momenta $k$. To better compare both information measures, we overprint the maximum IPR values at $\pm k_m\simeq \pm 0.229\,\mathrm{nm}^{-1}$ for valence (blue) and at $\pm k_m\simeq \pm 0.01\,\mathrm{nm}^{-1}$ for conduction (red) states, respectively. We also mark the momentum cutt-offs $k_c\simeq 0.383\,\mathrm{nm}^{-1}$ for valence (blue) and $k_c\simeq 0.03\,\mathrm{nm}^{-1}$ for conduction (red) first near-gap states obtained from the IPR criterion \eqref{cutoffdef}. In general, we could say that edge states participate of fewer momenta $k$ (higher IPR) than bulk states; they also have lower correlation modulus (but higher correlation arguments in absolute value) than bulk states for the real-space partition discussed here.

\section{Conclusions and outlook}\label{secconclu}

We have used QI theory concepts like IPR, RDM, entanglement entropies and space correlations, as  an interesting ``microscope'' to reveal  details of the internal structure of HgTe QW near-gap states with SOC (induced by the bulk and structural inversion asymmetries) in a finite strip geometry of width $L$, proposing a criterion for a momentum cutoff $k_c$  differentiating between edge- and bulk-states region behavior attending to IPR localization measures. To do this, we have considered a tight-binding Hamiltonian describing the low energy effective theory.  Quantitative information on the near-gap  energy spectrum and wave-functions is extracted from  a numerical Hamiltonian diagonalization approach, which is complemented by a previous analytic study of the uncoupled (without SOC) case. 

We corroborate previous results on the intriguing oscillatory dependence of the energy gap with $L$, this time for a more general SOC, with sudden gap drops for critical strip widths $L_c$. The non-trivial  consequences of the  Rashba term on the charge conductance are also reviewed, with a possible design of a QSH FET. 

The spin-polarization structure of edge states in position $y$ and momentum $k_x$ has also been evidenced by using probability density and IPR calculations and plots. The IPR analysis as a function of the momentum reveals that less and less momenta participate for near-gap states below the cutoff $k_c$ (edge-state region), when near-gap states start being localized at the edges $y=0,L$. 

Complementary information on the spin-polarization structure of near-gap states in $(k_x,y)$ space is extracted from the RDM to the spin subsystem.  Contour plots of the RDM entries show the extremal values of spin up and down and spin transfer probabilities in $(k_x,y)$ space.  Our findings indicate that the spin-polarization pattern, as shown by the diagonal elements of the RDM, is typical of the edge-state region  $|k|< k_c$, becoming less distinct as we approach the bulk-state region $|k|> k_c$. Another possible criterion for differentiating edge-state from bulk-state behaviors  could be this. Additionally,  a study of the spin-transfer probability amplitudes (RDM off-diagonal components/coherences) concludes that, in the edge-state region  $|k|< k_c$ (with the possible exception of a neighborhood of the $\Gamma$ point), the spin-transfer probability is significantly lower than in the bulk-state region $|k|> k_c$. Therefore, edge currents are more effective at preserving spin than bulk currents, one might argue.

Entropies of the RDM inform on  regions in $(k_x,y)$ space where the spin sector is highly entangled with the rest of the system, due to spin-orbit coupling.  
Entanglement exhibits greater variability within the low-momentum $k$ region, where edge states are located, compared to the bulk state region (characterized by relatively low entropy uniformity). The behavior of the quantum correlations does not seem to depend on the particular entanglement measure used, as discussed in Section \ref{Schlienzsec} above.

Correlation functions for a real-space partition report that the edge-state region (that of higher IPR in $k$ space) has lower correlation modulus (but higher absolute correlation argument)  values than the bulk state region (that of lower IPR).

If we want to compare between QI tools to distinguish the edge from the bulk states, we would say that, whereas the RDM and linear entropy entanglement measures provide better information than IPR on the spin-polarization structure of the near-gap states in $(k_x,y)$ space, the IPR and space-correlation measures give a sharper distinction between the edge  and the bulk  state separating momentum region. These separating transition regions occur at different momentum cut-off values for valence and conduction bands, and they are also sensitive to the model paramenters and strip width.

Concerning future research directions, the incorporation of electron-electron interactions in the BHZ model, as already done in recent studies \cite{Mai2023,PhysRevB.107.155106}, would greatly increase the physical significance of our study. We also believe that the analysis of topological phases using QI concepts and tools opens up new prospects for the development of new quantum technologies. For example, other QI concepts like ``fidelity-susceptibility'' turn out to capture magic twist angles arising in twisted bilayer graphene \cite{CALIXTO2025116199}. Exploring these novel quantum phenomena, termed as ``twistronic'' \cite{Cao2018unconventional,PhysRevB.95.075420},  and others, could lead to significant opportunities for advancing quantum information devices and opening up pathways to more robust, adaptable, and efficient quantum technologies.

\section*{Acknowledgments}
 We thank the support of Spanish MICIU through the
project PID2022-138144NB-I00. OC was on sabbatical leave, from September 2023 to August 2024, at Granada University, Spain. OC thanks support from the program PASPA from DGAPA-UNAM. We would like to thank our Master's thesis student, Alejandro Moreno, for his help at specific points in the development of the article.

\bibliography{bibliography_1.bib}

\end{document}